# Architecture and Performance Models for QoS-Driven Effective Peering of Content Delivery Networks

Mukaddim Pathan and Rajkumar Buyya

**Gri**d Computing and **D**istributed **S**ystems (GRIDS) Laboratory
Department of Computer Science and Software Engineering
The University of Melbourne, Parkville, VIC 3010, Australia
*{apathan, raj}@csse.unimelb.edu.au*

## ABSTRACT

The proprietary nature of existing Content Delivery Networks (CDNs) means they are closed and do not naturally cooperate. A CDN is expected to provide high performance Internet content delivery through global coverage, which might be an obstacle for new CDN providers, as well as affecting commercial viability of existing ones. Finding ways for distinct CDNs to coordinate and cooperate with other CDNs is necessary to achieve better overall service, as perceived by end-users, at lower cost. In this paper, we present an architecture to support peering arrangements between CDNs, based on a Virtual Organization (VO) model. Our approach promotes peering among providers, while upholding user perceived performance. This is achieved through proper policy management of negotiated Service Level Agreements (SLAs) between peers. We also present a Quality of Service (QoS)-driven performance modeling approach for peering CDNs in order to predict the user perceived performance. We show that peering between CDNs upholds user perceived performance by satisfying the target QoS. The methodology presented in this paper provides CDNs a way to dynamically distribute user requests to other peers according to different request-redirection policies. The model-based approach helps an overloaded CDN to return to a normal state by offloading excess requests to the peers. It also assists in making concrete QoS guarantee for a CDN provider. Our approach endeavors to achieve scalability and resource sharing among CDNs through effective peering in a user transparent manner, thus evolving past the current landscape where *non-cooperative and distinct* CDNs exist.

# 1. INTRODUCTION

With the proliferation of the Internet, popular Web services often suffer congestion and bottlenecks due to large demands made on their services. Such a scenario may cause unmanageable levels of traffic flow (i.e. flash crowd [2]), resulting in many requests being lost. Replicating the same content or services over several mirrored Web servers strategically placed at various locations is a method commonly used by service providers to improve performance and scalability. The user is redirected to the nearest server and this approach helps to reduce network impact on the response time of the user requests. A Content Delivery Network (CDN) [25] is such a network of surrogate servers spanning the Internet for better delivery of content to end-users. They provide services that improve network performance by maximizing available bandwidth, improving accessibility and maintaining correctness through content replication. Thus, they offer fast and reliable applications and services by distributing content to *edge* servers located close to end-users [10].

Existing commercial CDNs are proprietary in nature. They are owned and operated by individual companies. Each of them has created its own closed delivery network, which is expensive to setup and maintain. Running a global CDN is even more costly, requiring an enormous amount of capital and labor. In addition, content providers typically subscribe to one CDN and thus can not use the resources of multiple CDNs at the same time. Furthermore, commercial CDNs make specific commitments to their customers by signing Service Level Agreements (SLAs) [6][31]. An SLA is a contract between the service provider and the customer to describe provider's commitment and to specify penalties if those commitments are not met. If a particular CDN provider is unable to provide Quality of Service (QoS) to end-user requests, it may result in SLA violation and end up costing the provider. Therefore, the requirements for providing high quality service through global coverage might be an obstacle for new providers, as well as affecting commercial viability of existing ones. It is evident from the major consolidation of the CDN market, down to a handful of key players, which has occurred in recent years. For example, a recent news outburst reveals that two commercial CDNs Akamai Technologies, Inc. and Netli, Inc. have signed a definitive agreement for Akamai to acquire Netli in a merger transaction[1].

To ensure QoS while serving end-user requests a CDN is required to either provide all necessary distributed computing and network infrastructure (thus massively over provisioning its resources during day-to-day operation), or to be able to harness external resources on demand to meet any unexpected resource shortfalls. Therefore, the objectives of providing high quality services, reducing expenses, and avoiding adverse business impact could be achieved by establishing peering arrangements between CDN providers [4]. This large Internet-wide cooperation can be termed as a 'peering arrangement' [26] or internetworking [17] between CDNs, where some CDN providers may team up at some point in time to share resources and form an alliance in order to respond to or exploit a particular niche [27]. It virtualizes multiple providers and provides flexible resource sharing and dynamic collaboration between autonomous individual CDNs. In such a system, a CDN serves user requests as long as the load can be handled by itself. If the load exceeds its capacity, the excess user requests are offloaded to the CDN network of the peers. Thus, an overloaded CDN can redirect a fraction of the incoming content requests. This approach also provides a means to avoid long-term (i.e. periodic traffic pattern during a particular Web event) or short term (i.e. flash-crowds) bottlenecks [26][27].

Such peering arrangements are appealing, since it allows individual providers to achieve greater scale and network reach cooperatively than they could otherwise attain individually. In such a cooperative multi-provider environment, users are redirected across distributed set of Web servers deployed by partnering CDNs as opposed to individual servers belonging to a single CDN. Moreover, limited information about response time or service cost is typically available from individual CDNs, and load balancing control is retained by an individual provider within its own Web servers. Therefore, request-redirections must occur over distributed sets of Web servers belonging to multiple CDN providers, without the benefit of the full information available, as in the single provider case.

The challenges in adopting a peering arrangement between CDNs include designing a system that virtualizes multiple providers and offloads end-user requests from the primary provider to peers based on cost, performance and load. In particular we identify the following key issues:

- *When to peer?* The circumstances under which a peering arrangement should be triggered. The initiating condition must consider expected and unexpected load increases.
- *How to peer?* The strategy taken to form a peering arrangement between multiple CDNs. Such a strategy must specify the interactions among entities and allow for divergent policies between peering CDNs.
- *Whom to peer with?* The decision making mechanism used for choosing CDNs to peer with. It includes predicting performance of the peers, working around issues of separate administration and limited information sharing between peering CDNs.
- *How to manage and enforce policies?* How policies are managed according to the negotiated SLAs. It includes deploying necessary policies and administering them in an effective way.

In this paper, we present a novel architecture of a Virtual Organization (VO) [21] based system for forming peering CDNs. In our architecture, a CDN serves end-user requests as long as the load can be handled by itself. If the load exceeds its capacity, the excess end-user requests are offloaded to the Web servers of the peers. The VO-based peering system endeavors to cut expenses, improving locality while preserving user perceived performance at a satisfactory level. In this regard, we present an approach to perform QoS-driven modeling of the peering CDNs based on the fundamentals of queuing theory. We also demonstrate the performance comparison of four request-redirection policies within the peering CDNs model. Our aim is to show that the cooperation between CDNs through a peering arrangement upholds user perceived performance by providing target QoS according to SLAs. Our performance models can be used to reveal the effects of peering and to predict end-user perceived performance. Our approach endeavors to assist in making concrete QoS guarantees by a CDN provider. The main contributions of this paper are:

- an architecture for an open and decentralized system that supports the QoS-driven effective peering of CDNs;
- a policy-based framework for SLA negotiation among peering CDNs to ensure that requests are effectively served, meeting user QoS requirements;
- performance models to demonstrate the effects of peering and to predict user perceived performance;
- systematic performance analysis and measurement-based methodology to study the impact of key performance parameters such as server load and measurement errors that can be expected in a realistic system;
- an approach to measure the QoS level of a CDN provider to ensure it provides efficient services; and
- performance comparison of four request-redirection policies within the peering CDNs system model.

---

[1] Akamai Technologies, Inc., "Akamai to Acquire Netli," Press Release, February 5, 2007
http://www.akamai.com/html/about/press/releases/2007/press_020507.html.

The rest of the paper is structured as follows: Section 2 presents the related work. Our architecture for peering CDNs with a broad view of the VO creation steps and VO formation scenarios is described in Section 3. The next section enlightens the QoS aspect in peering CDNs. It is followed by the details on policy management in peering CDNs environment, with a particular focus on negotiated SLAs and defined policy levels. Section 6 outlines the performance models and our approach for measuring QoS performance of a CDN provider. Results are demonstrated in Section 7, followed by the decisive evaluation of our approach in Section 8. Finally, Section 9 concludes the paper with a brief summary of expected contributions and future directions.

## 2. RELATED WORK

Peering of CDNs is gaining popularity among researchers of the scientific community. Several projects are being conducted for finding ways to peer CDNs for better overall performance. Analyses of previous research efforts reveal a deficient progress to define the frameworks and policies for CDN peering. The reasons for this lack of progress are due to the complexity of the technological problems, legal and commercial operational issues that need to be solved in practical context.

In the following, we outline some of the related research efforts found in literature. They can be separated into different groups: architectural models vs. analytical performance models, and provider vs. end-user side perspective to assist peering. We start with investigating the architectures for enabling the peering concept and then present the related performance modeling approaches. We also mention the initiatives in this context from user-side perspective along with the deployed Peer-to-Peer (P2P) CDNs.

IETF has taken the first initiative to propose a Content Distribution Internetworking (CDI) Model [17]. It allows CDNs to have a means of affiliating their delivery and distribution infrastructure with other CDNs who have content to distribute. The CDI Internet draft assumes a federation of CDNs but it is not clear how this federation is built and by what relationships it is characterized. It recommends providing QoS in the cooperative domain either through using a supervision function or an independent third party to supervise and manage all the CDN peers, and to give some guarantees on the QoS of each CDN in the CDI model. However, it does not provide any hint on the type and/or characteristics of the supervision function to be used. The CDI model also does not examine the implications of using an independent third party for ensuring QoS guarantees to end-user requests. Based on the CDI model, Turrini [30] presents a protocol architecture, where performance data is interchanged between CDNs before forwarding a request by an authoritative CDN (for a particular group). This approach adds an overhead on the response time perceived by the users. Moreover, being a point-to-point protocol, if one end-point is down the connection remains interrupted until that end-point is restored. Since no evaluation is provided for performance data interchange, the effectiveness of the protocol is unclear.

CDN brokering [5] can be identified as a pioneering research effort towards developing a CDN brokerage system deployed on the Internet on a provisional basis. It allows one CDN to intelligently redirect end-users dynamically to other CDNs in a domain. The drawback is that the routing mechanism used is proprietary in nature and might not be suitable for a generic CDI architecture. Though it provides benefits of increased CDN capacity, reduced cost and better fault tolerance, it does not consider the end-user perceived performance to satisfy QoS while serving requests. Moreover, it only demonstrates the usefulness of brokering rather than to comprehensively evaluate a specific CDN's performance.

Cardellini et al. [11] presents an architecture for enhancing QoS in geographically distributed Web systems. It integrates DNS proximity and dispatcher scheduling with an HTTP redirection mechanism in order to achieve a scalable and balanced Web system. Though it aims to minimize the response time experienced by users while accessing geographically distributed Web sites, the use of HTTP request redirection may lead to increased network impact on latency experienced by the end-users. Moreover, it does not capture the heterogeneity in Web server systems since it only considers homogeneous Web clusters (hence, service distributions are the same) belonging to a single entity.

While the above mentioned research efforts do not explicitly virtualize multiple CDN providers, a peering system for content delivery workloads in a federated, multi-provider infrastructure is presented by Amini et al. [4]. The core component of the system is a peering algorithm that directs client requests to partner providers to minimize cost and improve performance. But the peering strategy, resource provisioning and performance guarantees among partnering CDNs are unexplored in this work.

In the context of modeling traffic redirection between geographically distributed Web servers, an approach to model the traffic redirection in geographically diverse server sets [3] and *W*ide *A*rea *R*edirection of *D*ynamic Content (WARD) [29] can be mentioned. The first uses a novel metric Server Set Distance (SSD) to simplify the modeling and classification of redirection schemes. Though this model provides a foundation for intelligent server selection over multiple, separately administrated server pools, it does not try to show the effectiveness of any particular policy or evaluate the QoS performance of any given CDN. On the other hand, WARD presents a novel architecture for redirecting dynamic content requests from an overloaded Internet Data Center (IDC) to a remote replica. It demonstrates a simple analytical model to characterize the effects of wide area request redirection on end-to-end delay. WARD can avoid over-provisioning of IDCs and achieve significant performance improvement through reduction in average request response times. However, it is mainly targeted to IDCs under the control of single administrative entity. Therefore, it does not virtualize multiple providers while considering request redirection. Moreover, it also does not provide mechanisms to evaluate the QoS performance of individual IDCs.

From a user-side perspective, Cooperative Networking (CoopNet) [23] addresses the flash crowd problem through the cooperation of end-hosts. CoopNet is found to be effective for small Web sites with limited resources. The main problem of the user-side mechanisms is that they are not transparent to end-users, which is likely to restrict their widespread deployment. Hence, it can not be used as a replacement and/or alternative for cooperation among infrastructure-based CDNs.

Among the deployed P2P-based CDN systems, CoDeeN [32], CoralCDN [19], Globule [28] can be mentioned as they provide collaborative content delivery. Other systems such as DotSlash [34] is a community driven "mutual" aid service that offers support for small sites that would not have the resources to cope during instances of flash crowds. Although CoDeeN and CoralCDN are built on top of a set of cooperative Web caches, they do not capture the notion of multi-provider existence for content delivery. On the other hand, Globule aims to allow Web content providers, not CDN providers, to organize together and operate their own world-wide hosting platform. While CoralCDN gives better performance to most users for accessing participating Websites, CoDeeN provides participating users better performance to most Web sites. However, none of them provide mechanism to evaluate the QoS performance of a certain provider. Moreover, some of these systems make strong assumptions on the characteristics of applications and do not virtualize multiple providers for cooperative management and delivery of content in a peering environment.

# 3. VO-BASED PEERING CDNs

A CDN is expected to provide the necessary distributed computing and network infrastructure to ensure SLAs are met with its customers. In order to meet such SLAs and to manage its resources properly, it could be necessary to cooperate with other CDNs through establishing peering arrangements between themselves. In this paper, we define a 'peering arrangement' between CDNs as:

> **Definition of 'peering arrangement' –** *A peering arrangement of CDNs is formed by a set of autonomous CDNs {$CDN_1$, $CDN_2$, …, $CDN_n$}, which cooperate through a mechanism **M** that provides facilities and infrastructure for cooperation between multiple providers for sharing resources in order to ensure efficient service delivery. Each $CDN_i$ is connected to other peers through a 'conduit' $C_i$, which assists in discovering external resources that can be harnessed from other CDNs. We denote **S** = {$S_1$, $S_2$, …, $S_m$} as the set of services provided by a CDN.*

Our definition of 'service' in this context is in line with the service definition in Service Oriented Computing (SOC) paradigm [24]. We define a 'service' as:

> **Definition of 'service'** – *A service $S_i$, offered by a cooperating $CDN_i$ is the endpoint of a well-defined, self-contained connection or underlying system (which does not depend on the context or state of other services) to serve a request according to the QoS requirements of end-users. A service request may specify a call for serving request for an individual file or object, a Web page (containing multiple objects) or an application service of a particular script (e.g. CGI, PHP) or any digital content.*

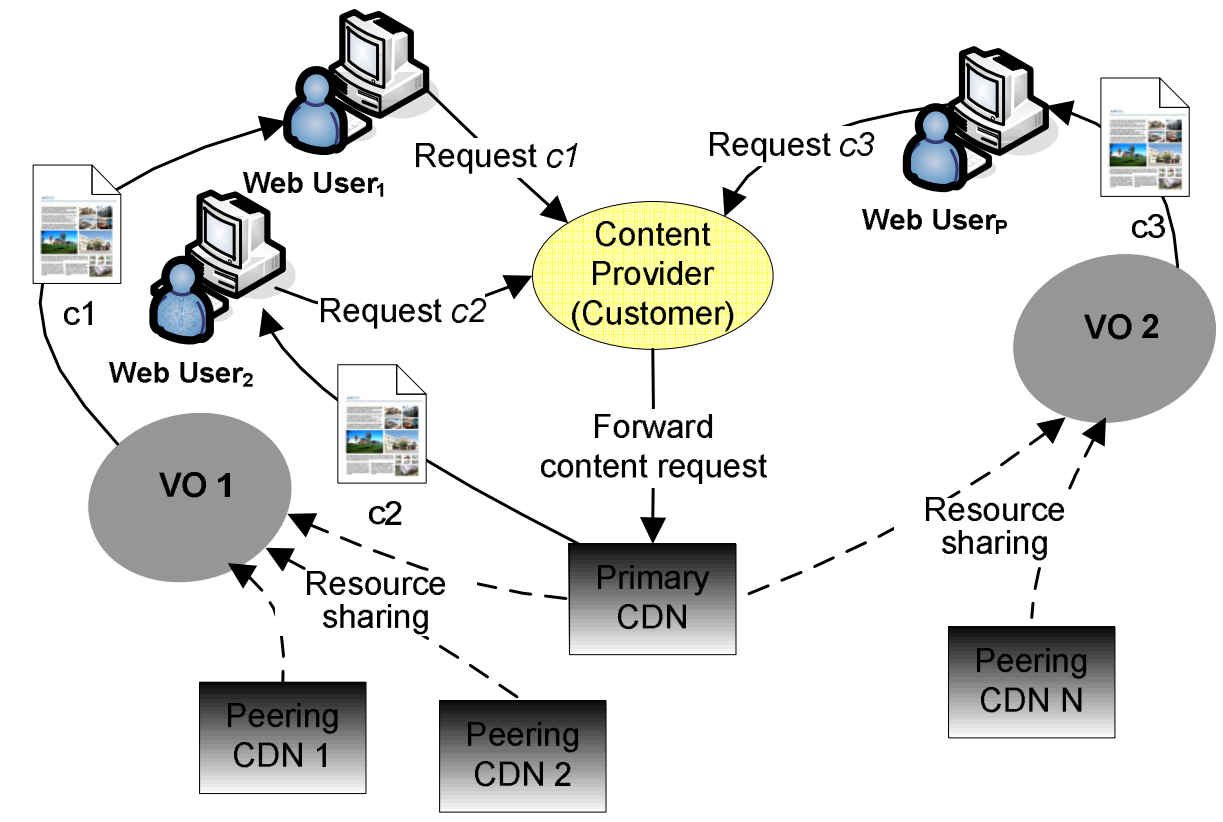

**Figure 1: Example of VO-based peering CDNs**

In our approach, cooperation among the peering CDNs is achieved through a VO. Here, we adopt the following definition of VO:

> **Definition of 'VO'** – *A VO is composed of a number of semi-independent autonomous entities (including different individuals, departments, and organizations) that come together to share resources and collaborate on common goal(s).*

A VO in the peering CDNs architecture consists of multiple autonomous CDNs who collaborate through a 'resource sharing' approach to serve user requests efficiently according to QoS requirements. Formation of a VO is initiated by a CDN, which realizes that it will not be able to meet its SLAs with the customers. The initiator is called a *primary* CDN; while other CDNs who share their resources in a VO are called *peering* CDNs. Users interact transparently with the VO by requesting content from Web servers of the primary CDN. A content request may initiate further VO activities that the end-users are unaware of (e.g. inter-CDN request-routing, content replication and delivery in a peering arrangement). Thus, the participating entities act as a single conceptual unit in the execution of common goal(s).

A VO is composed of *explicit* members who are the primary and any peering CDNs who cooperate for resource sharing, and *implicit* members who are content providers and end-users. Implicit members are transparent to a VO but they share the benefit from it. Figure 1 shows the example of VO-based peering CDNs. End-user requests for content are forwarded to the primary CDN which is holding the content from the content provider. The requested content is served either directly by the primary CDN or by any peering CDNs within a VO. Let us consider content c1 and c3 in Figure 1. Since c1 and c3 reside in the Web servers within VO1 and VO2 respectively, requests of c1 and c3 are served accordingly from VO1 and VO2. In case of content c2, the primary CDN directly delivers the requested content.

**Table 1: List of commonly used terms**

| Terminology | Description |
|---|---|
| *Web server (WS)* | A container of content |
| *Mediator* | A policy-driven entity, authoritative for policy negotiation and management |
| *Service registry (SR)* | Discovers and stores resource and policy information in the local domain |
| *Peering Agent (PA)* | A resource discovery module in the peering CDNs environment |
| *Policy repository (PR)* | A storage of Web server, mediator and VO policies |
| $P_{WS}$ | A set of Web server-specific rules for content storage and management |
| $P_M$ | A set of mediator-specific rules for interaction and negotiation |
| $P_{VO}$ | A set of VO-specific rules for creation and growth of the VO |

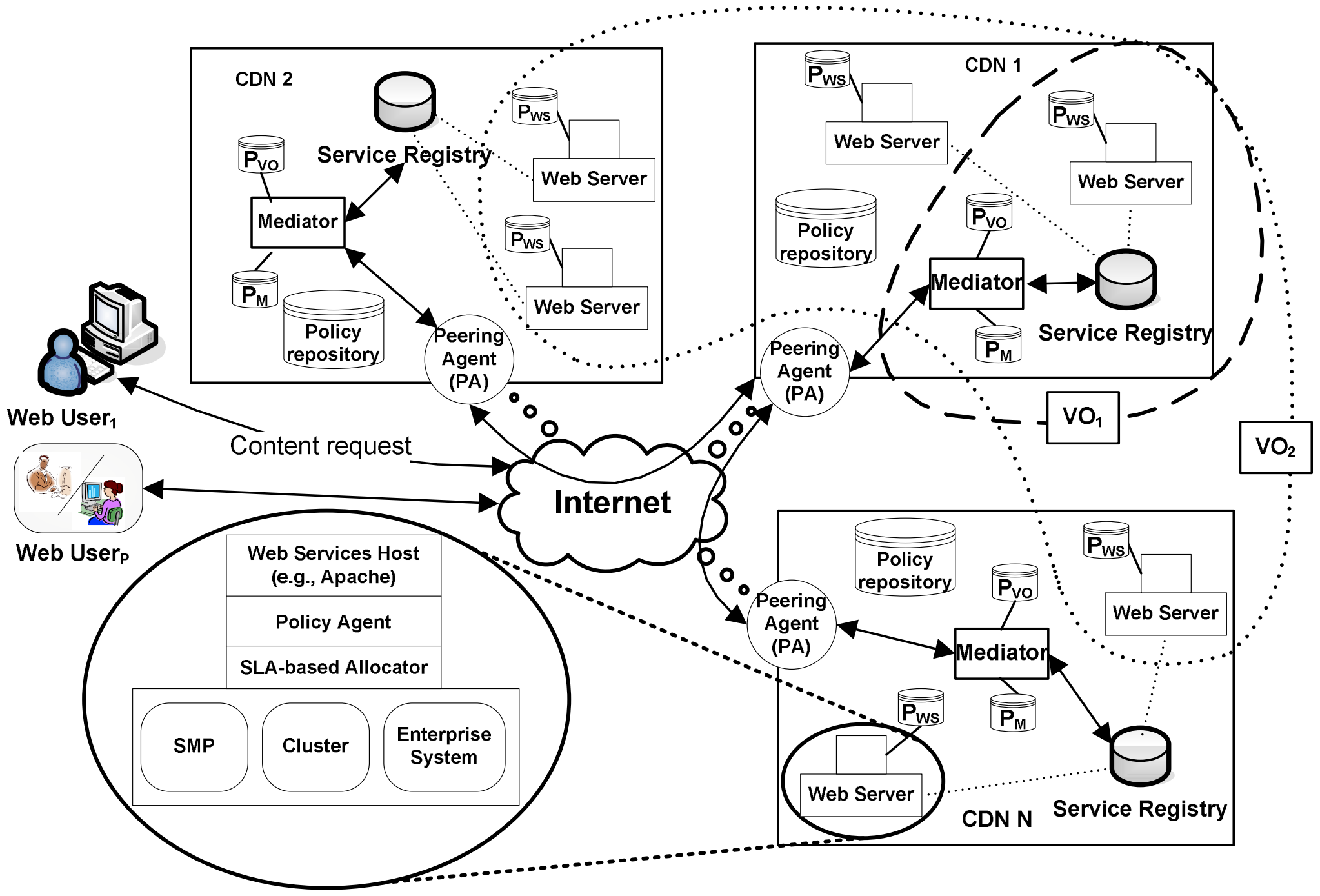

**Figure 2: Architecture of a system to assist the creation of peering CDNs**

## 3.1 System architecture

The architecture of our VO-based peering CDNs is shown in Figure 2. The terminologies used to describe the system architecture are listed in Table 1. In the VO-based peering CDNs model, a CDN endeavors to balance its service requirements against the high costs of deploying customer-dedicated, over-provisioned resources. Thus, to cut expenses and avoid the potential peak load threat of violating SLAs with the customers, CDNs will be able to leverage computing and network infrastructure from other CDNs through peering. The negotiation among CDNs for resource sharing allows a *peering* CDN to agree to allocate the required amount of its local resources (Web servers, bandwidth etc.) in order to provide content and services on behalf of the *primary* CDN.

A peering arrangement among CDNs that allows provisioning and sharing of computing resources must also provide settlement and exchange of the generated revenue. The primary CDN ultimately controls the resources it has acquired – which are delegated rights for the peering CDNs' physical resources. The physical resources could consist of resources from multiple peering CDNs distributed over different geographical locations. The primary CDN determines what proportion of the Web traffic (i.e. user requests) is redirected to the peering CDNs, what content is replicated, how the replication decisions are taken, and which replication policies are being used.

In our architecture, *Web Servers (WS)* within a CDN are the actual holders of content, comprising of two layers—overlay, which is a collection of Web service host (e.g. Apache, Tomcat), Service Level Agreement (SLA)-allocator, and policy agent; and core, which refers to the underlying hardware infrastructure. Each Web server has its own policies, defined as a set of server-specific rules, $P_{WS}$, for the storage and management of content. The *Service Registry (SR)* helps in discovering local resources within a CDN by providing resource and access related information. The *Peering Agent (PA)*, *Mediator* and *Policy Repository (PR)* collectively act as a "gateway" for a given CDN, and all three assist in creation of a new VO. The PA of a CDN acts the role of a resource discovery module. It acts as the first point of contact for other CDNs when they are initiating a peering agreement, and a conduit through which a CDN can itself discover potentially useful resources available from other CDNs. The mediator is responsible for negotiation among CDNs and management of operations within a VO. The mediator has its own policies (defined as a set of mediator-specific rules, $P_M$) and also manages the policies (defined as a set of VO-specific rules, $P_{VO}$) necessary for negotiation and creation of VO(s). The PR virtualizes all of the individual policies from within the VO, including $P_{WS}$, $P_M$, and $P_{VO}$, and will ultimately include policies from peering CDNs.

## 3.2 Lifecycle of a VO

A VO may vary in terms of purpose, scope, size, and duration. Hence, VOs are of two types: *short-term on-demand VOs* and *long-term VOs* with established SLAs. A short-term VO is formed for limited duration, based on the current user request patterns to prevent the generation of *hotspots* [2]. Such a peering arrangement should be automated to react within a tight time frame – as it is unlikely that a human directed negotiation would occur quickly enough to satisfy the evolved niche. A short-term VO is formed on-demand and the policy for such VO formation is established dynamically to handle the situation, one such negotiation mechanism is described in section 3.3. Short-term VOs are phased out when the workload returns to normal. On the other hand, a long-term VO is formed for an event which will be known in advance. In a long-term VO, CDNs collaborate for longer period of time and such a VO remains for the duration of the event. In this situation, we would expect negotiation to include a human-directed agent to ensure that any resulting decisions comply with participating companies' strategic goals. Relevant scenarios for short and long-term VO creation have been illustrated in [27].

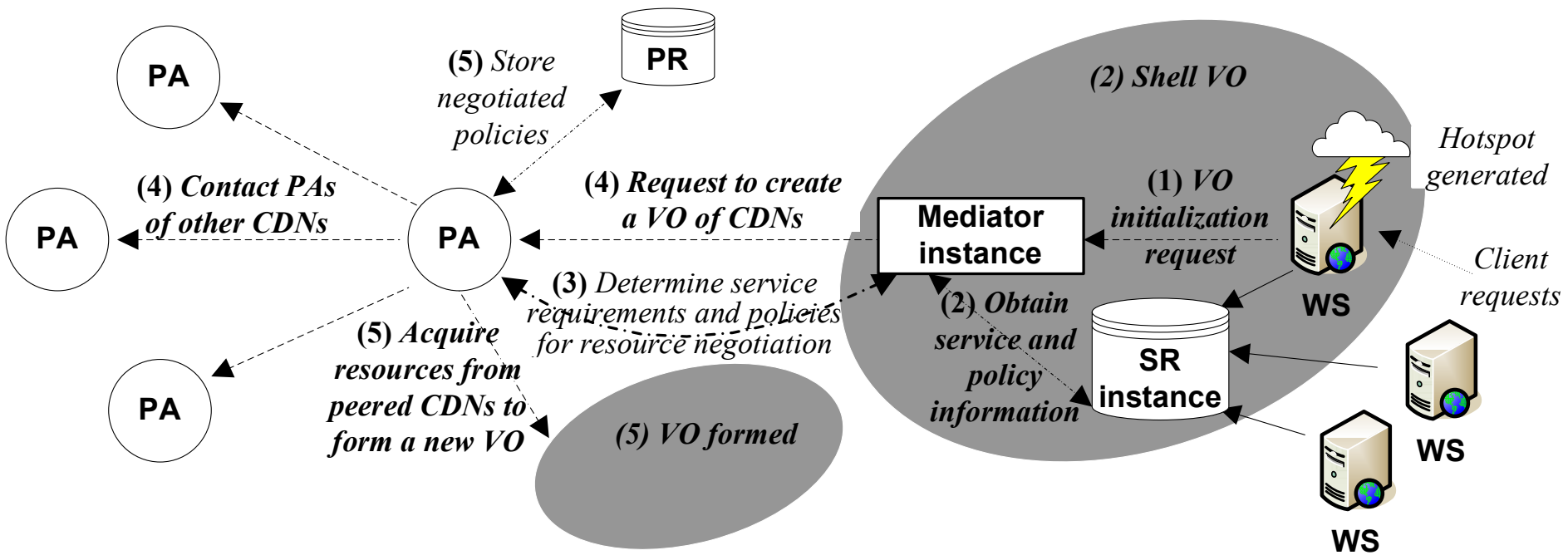

**Figure 3: The formation of a Virtual Organization (VO)**

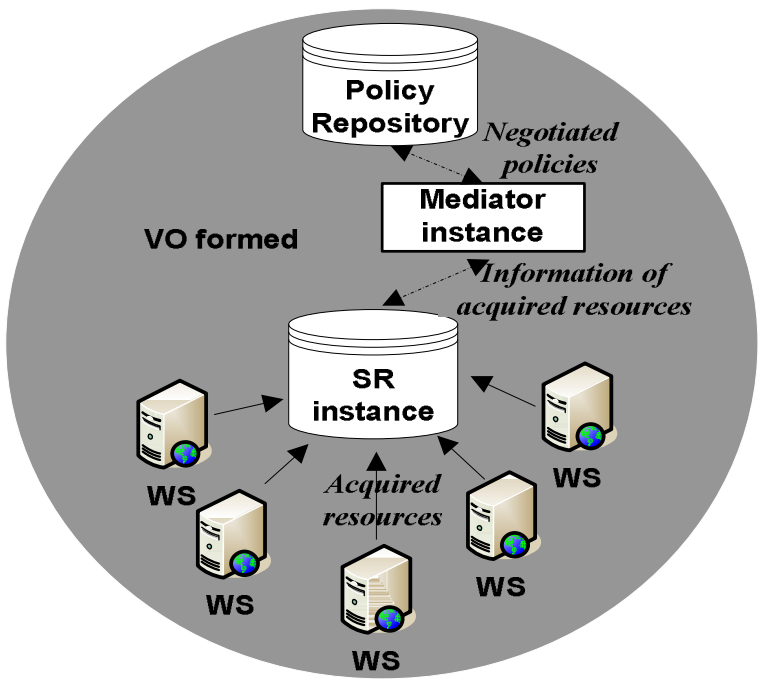

**Figure 4: A formed VO**

Figure 3 and Figure 4 respectively illustrate the VO creation steps and a VO after it is formed, while Figure 5 shows the flowchart of a VO lifecycle. When a VO is formed, a number of issues are taken into account, including:

- An entity responsible for forming a VO should be able to recognize circumstances in which it should initiate VO formation.
- A participating entity while joining in a VO should be able to determine the conditions under which it is profitable for it to join.
- Given a number of offers, the entity that initiates VO formation should be able to determine the best offer(s) in terms of economic benefit

Figure 6 presents the protocol for cooperation between CDNs through VO formation. The major phases of the protocol are the following:

**Phase I: Hotspot detections** – A (primary) CDN provider realizes that it cannot handle a part of the workload on its Web server(s). A VO initialization request is sent to the mediator.

**Phase II: Shell VO creation –** The primary CDN constructs a *shell VO*, with a mediator instance, a service registry instance, and a policy registry. The mediator instance obtains the resource and access information from the SR, whilst SLAs and other policies from the PR

**Phase III: Expansion of shell VO –** The shell VO represents the potential for peering of resources. Hence, it needs to be expanded to include additional resources from other CDNs. The mediator instance, on the primary CDN's behalf, generates its service requirements based on the current circumstance and SLA requirements of its customer(s). The mediator instance passes the service requirements to the local Peering Agent (PA). If there are any preexisting peering arrangements (for a long term scenario) then these will be returned at this point. Otherwise, short term negotiation(s) is carried out between the PA identified peering targets (Section 3.3).

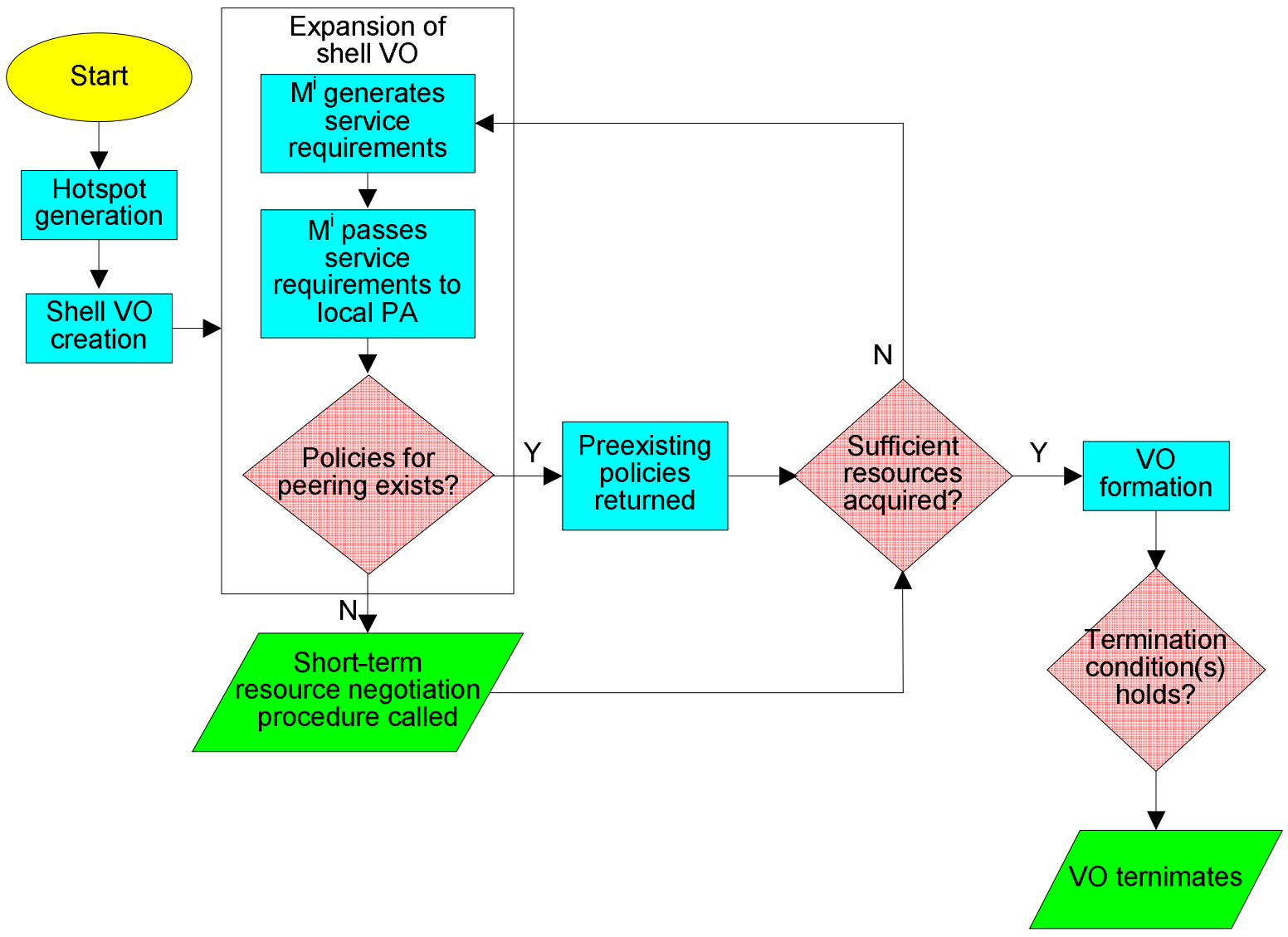

**Figure 5: Flowchart of a VO lifecycle**

---

**Algorithm** Cooperation between multiple CDNs

---

**Input:** $N = \{1, 2, 3, \ldots, N\}$ CDNs;
**Output:** $VO = \{VO_1, VO_2, \ldots\}$: Virtual Organization(s) of Peered CDNs
**Phase I: Hotspot detections**
1: Primary CDN cannot handle a part of the workload, $RT(r_i, t_k) > D$
2: Mediator receives a VO initialization request
**Phase II: Shell VO creation**
3: Primary CDN constructs a *Shell VO*
*Shell VO* ← $\{M^i, SR^i, PR\}$ /* $M^i$: Mediator instance, $SR^i$: Service Registry instance, *PR*: Policy Repository */
**Phase III: Expansion of Shell VO**
4: *Shell VO* represents the potential for peering of resources
5: $M^i$ generates *Service Requirements*, $S^R$
$S^R \leftarrow \{S_c, D, P, T\}$ /* $S_c$: Capacity Requirements, *D*: Delay threshold, *P*: Preference, *T*: Duration */
6: $M^i$ passes service requirements to local *PA*
7: **if** ($E_{policy} \neq$ NULL) **then** /* $E_{policy}$: Preexisting policies */
8: **return** $E_{policy}$
9: **else**
10: $P^* \leftarrow$ **Procedure(S^R)** /*$P^*$: set of peers who meet $S^R$ */ (*Section 3.4*)
11: **end-if**
**Phase IV: VO formation**
12: **if** ($P^* \neq \{\phi\}$) **then**
13: Primary CDN acquires sufficient resources from the peers
14: VO is formed and it becomes operational
15: **else**
16: Renegotiation is resumed from step 3 with reconsidered $S^R$
17: **end-if**
**Phase V: VO termination**
18: **for** *i* = 1 to 4 **do**
19: **if** ($D^c[i]$ = TRUE) **then** /* $D^c[i]$: set of conditions = {the circumstances under which the VO was formed no longer hold, peering is no longer beneficial for the participating CDNs, an existing VO needs to be expanded to deal with additional load, participating CDNs are not meeting the agreed upon contributions}
20: Disband or re-arrange VO
21: **end-if**
22: **end for**

---

**Figure 6: Protocol for cooperation between CDNs**

.**Phase IV: VO formation –** When the primary CDN acquires sufficient resources from its peers to meet its SLAs with the customer, the new VO becomes operational. If no CDN is interested in such peering, VO creation through re-negotiation is resumed from Step 3 (Phase II) with reconsidered service requirements.

**Phase V: VO termination –** An existing VO may need to either disband or re-arrange itself if any of the following conditions hold: (a) the circumstances under which the VO was formed no longer hold; (b) peering is no longer beneficial for the participating CDNs; (c) an existing VO needs to be expanded further in order to deal with additional load; or (d) participating CDNs are not meeting their agreed upon contributions.

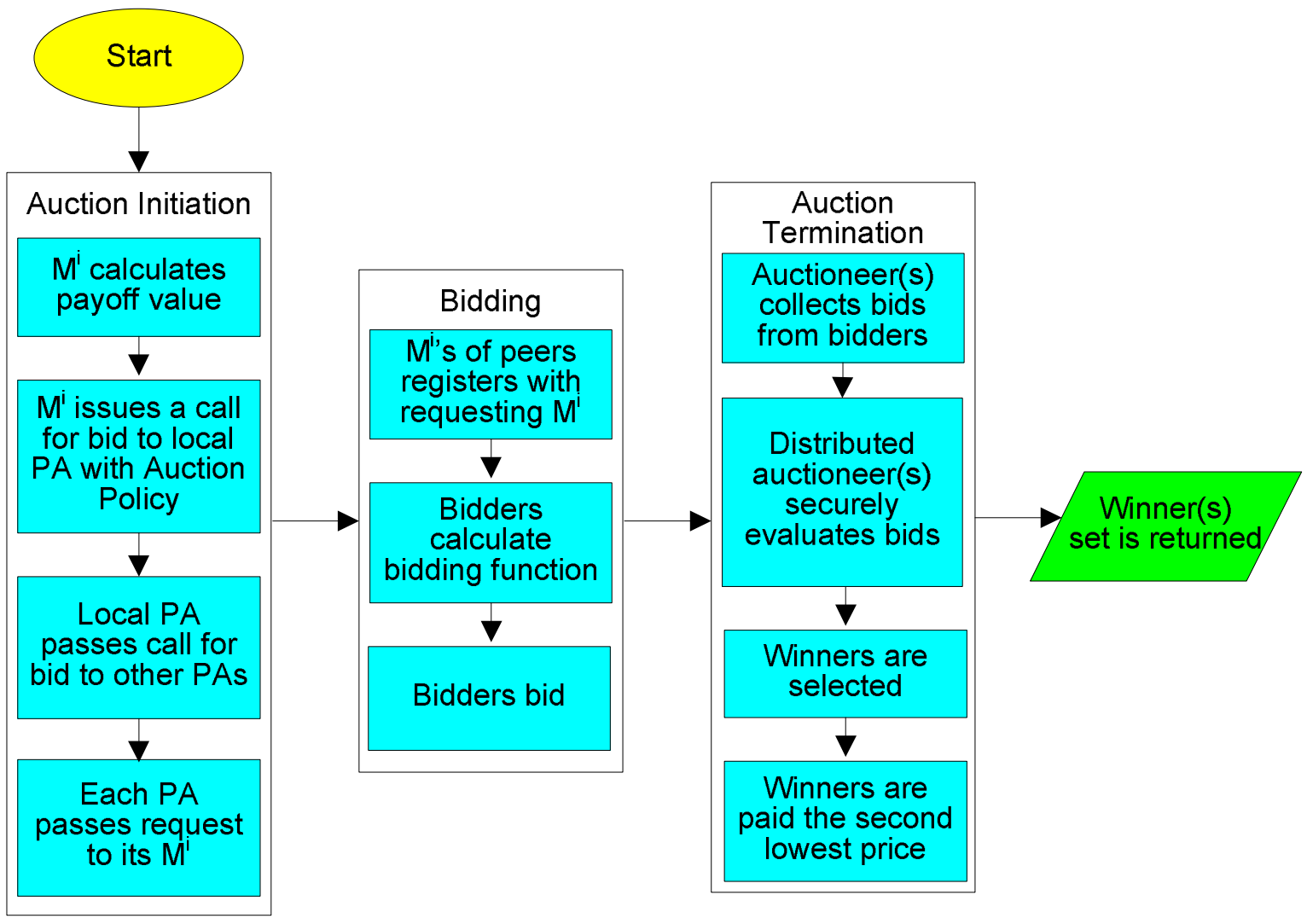

**Figure 7: Flowchart for short-term resource negotiation through auction**

---

**Procedure($S^R$):** Autonomic Resource Negotiation

---

**Input:** $N$: set of CDNs, $C$: set of content, $r_k = \{r_k^c, r_k^l\}$: k-th content request

**Output:** Peering CDNs to participate in a cooperation

**Phase I: Auction initiation**

1: $M^i$ calculates *Payoff Value*, $P_{max}$
   $P_{max} \leftarrow$ (the managing cost for $r_k^c$) + (the expected profit from $r_k^c$)

2: $M^i$ issues a call for bid to local *PA*, with *Auction Policy*, $A_p$
   $A_p \leftarrow \{S_c, D, P, T\}$ /* $S_c$: Capacity Requirements, $D$: Delay threshold, $P$: Preference, $T$: Duration */

3: *PA* distributes auction request to other *PA*s of the peers

4: subset of $M^{i'}$s are selected as the distributed *Auction Evaluators*

5: Each *PA* passes the request to its $M^i$

**Phase II: Bidding**

6: $M^i$'s of the peers register with the requesting $M^i$

7: Bidders calculate *Bidding Function*, $B_i(r_k)$

$$B_i(r_k) = S_i(r_k^c) + ER_i(r_k^c, t_k, n) + \psi_i(A_P)$$ /* $S_i(r_k^c)$: Cost incurred,

$ER_i(r_k^c, t_k, n)$ : Expected Revenue, $\psi_i(A_P)$: Seller $i$'s interest to bid */

$$ER(r_k^c, t_k, n) = \alpha \sum_{j=k+1}^{k+n} \delta(r_k^c, r_j^c)\phi(r_k^l, r_j^l) + (1-\alpha)\sum_{i=1}^{k-1} \delta(r_k^c, r_i^c)\phi(r_k^l, r_i^l), 0 \le \alpha \le 1$$

/* $\delta(c_i, c_j)$: Similarity function for two content, $0 \le \delta(c_i, c_j) \le 1$; $\phi(r_i^l, r_j^l)$: Similarity of two content requests in terms of distance, $0 \le \phi(r_i^l, r_j^l) \le 1$

**Phase III: Auction Termination**

8: Auctioneer collects bids from bidders

9: Bids are evaluated securely by distributed auction evaluators

10: Winners' set, $W$ is selected based on lowest bid

11: Winner(s) are paid the amount of second-lowest bid

12: **return** $W$

---

**Figure 8: Auction protocol for short-term resource negotiation**

## 3.3 Short-term resource negotiation

In order to respond to hotspots that may result in a CDN failing to meet its QoS obligations, we propose that time-critical agreements for a short-term VO should be automatically negotiated. We expect any such agreements to hold for a limited duration and only involve an artificially restricted set of CDN resources. Even so, there are serious issues involving such agreements including trust, i.e.

who governs the allocation decisions and are they trustworthy; and the potential commercial sensitivity of information about the current state and costs of a CDN. Divulging commercially sensitive information (e.g. resources, access and policies) as a basis for negotiating a peering arrangement would be, in general, commercially unacceptable. Even with the limitations placed on resources that can be automatically delegated, there must be checks and balances to ensure that any delegation is made properly. Otherwise, it would also be unlikely that a CDN would agree to have an external party (e.g. mediator of the primary CDN) make allocations of their resources and bind them to negotiated SLAs.

One solution to these problems is to utilize a cryptographically secure auction [8], which hides both the valuations that CDNs place on their resources and who is participating in the auction. With this approach, no CDN needs to trust a particular auctioneer; rather the auction mechanism itself is trustworthy due to the cryptographic guarantees. Thus, malicious behavior by any CDN (including the primary) or third party auctioneer is prevented and sensitive bid prices are kept secret. These auctions have been shown to be tractable in practice and are therefore an ideal basis for automation of peering agreements.

A negotiation would start with the VO being formed and the mediator determining the shortfall in resources. The mediator then issues its local PA a call for bids and within this includes the SLAs that it requires. The PA distributes the request to other PAs of the peers and each of them then passes the request to its CDN mediator. All mediators that wish to bid then register with the requesting mediator and a subset of the mediators (acting as the *distributed auctioneers*) are selected via a cut and shuffle [22]. Note that the requesting mediator does not act as an auctioneer. The auction is then held securely and only the final resource allocations are revealed. The auction is combinatorial as multiple resource types can be specified and therefore different groupings and even substitutions of those resources over the peering CDNs are possible.

An auctioneer starts an auction not for selling an item (i.e. allocation), but for buying it. Buyers (peering CDNs) bid with the price they are willing to sell the allocation of their Web servers. One bidder can not see the bid of other bidders. An auctioneer gathers bids from the bidders and selects the lowest bidding agent(s) as the winner and the winner is paid second-lowest bidding price. In other words, a reverse Vickrey auction is used. As mentioned earlier, we assume that an auction is held using a cryptographically secure auction [8] protocol to hide all auction related sensitive information. Through this approach, a mendacious behavior from a provider is restricted. Thus, over-provisioning of resources by harnessing data through VO membership, or modifying and falsifying of content by some rogue CDN providers is not allowed.

Figure 7 shows the flowchart for autonomic resource negotiation through auction. Figure 8 presents the corresponding auction protocol. Here, we also summarize the steps for the auction to be held within a VO:

**Phase I: Auction initiation –** Before the start of the auction a subset of the mediators from the potential peering CDNs are randomly selected to act as the distributed *Auctioneers*. Auction starts when the mediator of a CDN (buyer) realizes the need of additional resources to replicate content. It internally determines the maximum payable amount (expressed by *Payoff Value*). The mediator then issues its local PA a call for bids including its service requirements (*Auction Policy*).The PA distributes the bidding request to other PAs of the peers and each of them passes the request to its mediator.

**Phase II: Bidding –** All mediators that wish to bid then register with the requesting mediator. All bidders (peers) use a *Bidding Function* to determine the bidding amount and potential peering CDNs bid their price.

**Phase III: Auction termination and re-negotiation –** An auctioneer collects bids and these bids are then evaluated securely by the distributed auctioneers to determine the winners and the optimal combinations of resources. An auction takes place successfully when winners are chosen according to the Auction policy of the primary CDN. A winner is paid by the amount of second-lowest bid. At this point, it can be assumed that a CDN has acquired sufficient resources from its peers to meet its SLAs with customers. If no winner is selected, re-negotiation through auction takes place, starting from Phase I.

# 4. QoS IN PEERING CDNs

The definition of QoS relates to the agreements between a service provider and its customers. In the VO-based peering CDNs environment, such agreements are specified as SLAs, which results in a fixed set of well understood terms—in our case, the QoS requirements of end-users. The definition of quality varies from different perspective and views. The following three views of quality are the most common:

- *Quality as functionality* – According to this view, quality is considered in terms of the amount of functionality that a service provider can offer to its customers. Quality as functionality characterizes the design of a service and can only be measured by comparing the service against other services offering similar functionalities. For example, if CDN A provides content delivery service for static content only, while CDN B provides the same for both static and dynamic content; then CDN B can be considered to offer better quality than CDN A.

- *Quality as conformance* – In this view, quality is seen as being synonymous with meeting provider's commitments and specification such as SLAs. Quality as conformance, which can be monitored for each service individually, usually requires the user's experience of a service in order to measure the 'promise' against the 'delivery'. For example, if CDN A makes the commitment to its customer that 95% of the user requests will be served within less than 2 seconds, and it maintains it at all times in its operation, then CDN A is usually considered as offering good QoS.

- *Quality as reputation* – In this view, quality is linked to user's perception of a service in general. This perception is developed gradually over the time of a service provider's existence. Quality as reputation can be regarded as a reference to a service provider's consistency over time in offering both functionality and conformance qualities, and can therefore be measured through the other two types of qualities over time. It can also be viewed as the 'goodwill' of a particular service provider. For example, Akamai [1][18] is generally considered to be offering good quality content delivery services to the users, due to its reputation and large market share developed in course of time as a CDN provider.

While it is possible to consider all three types of quality for a service in the peering CDNs environment, it is perhaps most interesting and relevant to address the issue of ensuring QoS from the view of quality as conformance. Therefore, in this paper, we adopt the conformance view and define QoS as the experience perceived by a user for being served by a CDN. Specifically, in regard of the peering CDNs, we define quality in the following way:

**Definition of 'quality' –** *Let A be a CDN provider and* $\mathbf{S} = \{S_1, S_2, \ldots, S_m\}$ *be the set of services provided by it. Assume that for each service* $S_i$, $S_i^p$ *is the quality that A promised to offer to the users and* $S_i^d$ *is the actual quality delivered by A. Then the QoS for CDN A is given by,*

$$QoS_A = f(S_i^p, S_i^d)$$

*where f is the function that measures the conformance between* $S_i^p$ *and* $S_i^d$.

In this definition, we capture the notion of conformance generally, but do not specify how $S_i^p$ and $S_i^d$ can be measured. We anticipate measuring the QoS assessment of a CDN in terms of the *response time* perceived by a user when it receives services from it. We also state that the QoS measure may take into account not only the average response time, but also the percentile (95$^{th}$ percentile, for example) of the response time. We specify response time as the time a service takes to respond to various types of requests. It is a function of load intensity, which can be measured in terms of expected waiting time of a request to be served (Section 6).

## 4.1 SLAs to ensure QoS

Ensuring QoS guarantees requires a means of establishing a set of common quality parameters and establishing which attributes are needed by a particular customer to describe its QoS requirements. These factors are combined in an SLA that both a customer and provider agree to and that the provider refers to when monitoring its QoS performance. Examples of QoS parameters that an SLA may specify are:

- The maximum response time for a service request will not exceed 0.5 seconds.
- 95% of user requests will be completed in less than 2 seconds.
- A service will be available for at least 99.9% of the time.

From a service management and business perspective, the fulfillment and assurance of SLAs is of key importance. In our context, the importance of an SLA lies in its encoding of performance obligations, so that a CDN can manage its workload. For example, the existing SLAs that a CDN currently holds firstly permit it to determine if a new SLA can be accepted as it is. Secondly, the SLAs are used to determine if the CDN is meeting user QoS requirements. Thirdly, a CDN uses its SLAs to quantify what further resources it would require in a peering arrangement. Thus, the existing SLAs are critical in establishing the runtime performance metrics for the CDN and as a basis for establishing peering arrangements. Examples of attributes that an SLA encodes are:

- *Service type* – What service (e.g. content and/or application delivery) a user is requesting from a CDN.
- *Service requirements* – The processing and/or capacity requirements for a CDN to serve user requests.
- *Duration* – The maximum time duration within which content requests are to be served from a CDN.
- *Guaranteed QoS level* – The level of guarantee (in terms of response time) that requests will be served within a delay threshold. For example, three levels of QoS guarantees can be specified: *Gold* = user requests will be served timely in 95% of cases; *Silver* = user requests will be served timely in 60% of cases; *Bronze* = no guarantee (best effort) will be provided for serving user requests.

# 5. SLA NEGOTIATION AND POLICY MANAGEMENT

When CDNs peer according to a VO-based framework, the participants sign SLAs with different performance objectives. Once the SLAs have been agreed upon, the participants in a VO work in order to satisfy the negotiated SLAs. The SLA components include:

- *Description of service requirements*, a specification of the resource and service requirements of the primary CDN. This description includes the storage requirements, the required rate of transfer, the primary CDN's preference to gain resources at a particular region, and the expected duration of receiving service.
- *Administration for VO activities*, which specifies the role of the mediator as an authoritative entity in the VO.
- *Renegotiation for problem resolution*, which illustrates the steps to be undertaken in face of any problem in providing necessary services.
- *Consequences of SLA violation*, which outlines the possible results of SLA violation in which service expectations are not met. The consequences of SLA violation may range from imposing penalty on the participants through reimbursement of part of the revenues lost due to the loss of service, to termination of peering relationship, and to disbanding and/or rearranging the participants to form a new VO.
- *SLA bypassing conditions*, which details the conditions under which the SLAs are not applicable. Such situations include the damage of physical resources due to natural disaster, theft etc.

## 5.1 Policy management to support SLAs

The proper operations of a VO-based peering CDNs architecture seek for the consistent performance and availability of a large number of widely distributed system entities, specified in an SLA. A policy-based framework can simplify the complexities involved in the operation and management of a large content distribution network [31]. Within our VO-based peering CDNs architecture we apply the standard policy framework defined by the IETF/DMTF [33]. The policy framework consists of four basic elements: *policy management*, *policy repository*, *policy enforcement point (PEP)*, and *policy decision point (PDP)*.

**Table 2: Policy mapping**

| Policy framework Component | Peering CDNs Component | Specified policies | Description |
|---|---|---|---|
| *System* | *Peering CDNs* | All policies in the system | The distributed computing and network infrastructure for peering CDNs |
| *Admin domain* | *Formed VO* | Negotiated VO policies | An administrative entity for resource management and access control |
| *Policy management tool* | *Administrator dependent* | – | An administrator dependent tool to generate policies |
| *Policy repository* | *Policy repository* | Web server, VO and mediator policies | Storage of policies in the system |
| *Policy Enforcement Points (PEPs)* | *Web Services host, Policy Agent, SLA-based allocator* | Web server policies | A logical entity which ensures proper enforcement of policies |
| *PDPs* | *Mediator* | Mediator policies, VO policies | An authoritative entity for retrieving policies from the repository |

The framework (and the corresponding entities) for peering CDNs can be mapped to the basic policy framework. We show this mapping in Table 2. The policy repository (PR) from Figure 2 virtualizes the Web server, mediator and VO policies. These policies are generated by the policy management tool used by the VO administrator. The distribution network and the Web server components (i.e. Web Services host, Policy Agent, SLA-based Allocator) are the instances of PEPs, which enforce the peering CDN policies stored in the repository. The mediator is the instance of the PDPs, which specifies the set of policies to be negotiated at the time of collaborating with other CDNs, and passes them to the peering agent at the time of negotiation. The policy management tool is administrator dependent and it is not shown in Figure 2.

We also define three levels of policies, namely—*Web server policy*, *mediator policy* and *VO policy*. These three policy levels are detailed in the following:

*Web server policies* specify,

- how a server performs consistent content caching;
- how the policy agent module operates based on negotiated SLAs; and
- how the SLA-based allocator module ensures provisioning of resources to satisfy negotiated SLAs.

*Mediator policies* specify,

- how the mediator interacts with the PA to pass information on service requirements;
- how the mediator takes over the administration of delegated resources once they are acquired;
- how the mediator effectively manages VO activities to cope with changing circumstances; and
- how the mediator coordinates with other entities to assist in resource allocation.

*VO policies* include,

- the policies necessary for initiating VO creation;
- the policies need to be administered in face of SLA violation by VO participants; and
- the policies to dynamically disband or rearrange a VO.

# 6. PERFORMANCE MODELING

The general objective of a peering CDNs model is to provide improved QoS performance through minimizing end-user response time. However, the proprietary nature of existing commercial CDNs makes it difficult to predict the performance that a given user could experience from a particular CDN. Furthermore, such a model can be based on a complex combination of attributes such as Web server responsiveness or load, expected network delay, or geographic location. Several of these potential attributes vary over time and there is no single repository for listing the value of attributes such as geographic location or expected delay for all Internet-connected systems. Therefore, the values used in a peering CDNs model are likely to be based on heuristics.

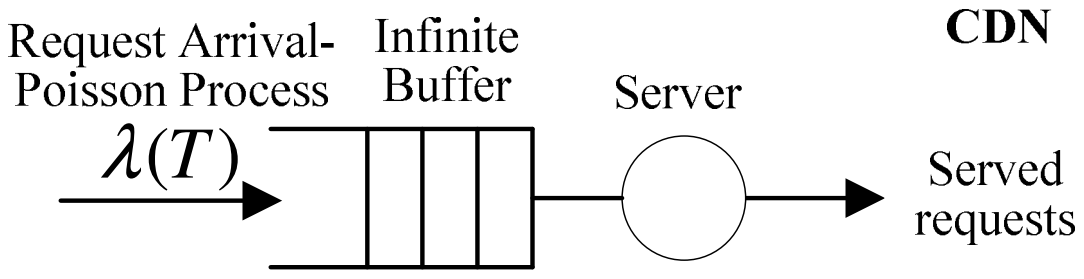

**Figure 9: Model of an M/G/1 queue**

**Table 3: Parameter and expressions for the analytical model**

| Parameter | Expression |
|---|---|
| *Mean arrival rate* | $\lambda = (1/T)$ (requests/second) |
| *Mean arrival time* | $T$ |
| *Mean service rate* | $\mu$ |
| *Load* | $\rho = (\lambda/\mu)$ |
| *P. D. F of service distribution* | $f(x) = \frac{\alpha k^{\alpha}}{1-(k/p)^{\alpha}} x^{-\alpha-1}, k \le x \le p$ |
| *Task size variation* | $\alpha$ |
| *Smallest possible task size* | $k$ |
| *Largest possible task size* | $P$ |
| *Expected waiting time* | $E[W]$ |
| *Mean service time* | $E[X]$ |

In this section, we develop simplified performance models based on the fundamentals of queuing theory to demonstrate the effects of peering between CDNs and to characterize the QoS performance of a CDN. For a given workload, mean and service time, we derive necessary expressions for measuring expected waiting time and cumulative distribution in order to measure the QoS performance of a CDN.

## 6.1 Single CDN model

Let us model a CDN as an M/G/1 queue as shown in Figure 9. Table 3 shows the parameters and expressions that are used in the modeling. The request streams coming to the Web servers of a CDN are abstracted as a single request stream. User requests arrive following a Poisson process with the mean arrival rate $\lambda$. All requests in its queue are served on a first-come-first-serve (FCFS) basis with mean service rate $\mu$. It is assumed that the total processing of the Web servers of a CDN is accumulated through the server and the service time follows a general distribution. The term 'task' is used as a generalization of a request arrival for service. We denote the processing requirements of an arrival as 'task size'. Here, we will use the terms task and request arrival interchangeably. A request can be a client requesting an individual file or object, a Web page (containing multiple objects), the results of execution of a script (e.g. CGI, PHP) or any digital content.

It has been observed that the workloads in Internet are heavy-tailed in nature [14][15], characterized by the function, $\Pr\{X > x\} \sim x^{-\alpha}$, where $0 \le \alpha \le 2$. In a CDN, clients request for content of varying sizes (ranging from small to large). Based on size of the content requested, the processing requirements (i.e. task size) also vary. Hence, we model the task size on a given CDN's service capacity to follow a Bounded Pareto distribution. The probability density function for the Bounded Pareto $B(k, p, \alpha)$ service distribution is

$$f(x) = \frac{\alpha k^{\alpha}}{1-(k/p)^{\alpha}} x^{-\alpha-1},$$

where $\alpha$ represents the task size variation, $k$ is the smallest possible task size, and $p$ is the largest possible task ($k \le x \le p$). By varying the value of $\alpha$, we can observe distributions that exhibit moderate variability ($\alpha \approx 2$) to high variability ($\alpha \approx 1$).

We start with the derivation of the expectation of waiting time $E[W]$, $W$ is the time a user has to wait for service. $E[N_q]$ is the number of waiting customers and $E[X]$ is the mean service time. By Little's law, the mean queue length $E[N_q]$ can be expressed in terms of the waiting time. Therefore, $E[N_q] = \lambda E[W]$ and load on the server, $\rho = \lambda E[X]$. Let $E[X^j]$ be the $j$-th moment of the service distribution of the *tasks*. We have,

$$E[X^j] = \begin{cases} \frac{\alpha p^j((k/p)^{\alpha} - (k/p)^j)}{(j-\alpha)(1-(k/p)^{\alpha})} & \text{if } j \ne \alpha \\ \frac{\alpha k^{\alpha} \ln(p/k)}{(1-(k/p)^{\alpha})} & \text{if } j = \alpha \end{cases}$$

Hence, using P-K formula, the expected waiting time $E[W] = \lambda E[X^2]/2(1-\rho)$. It can be used to measure the waiting time with respect to different server load and task sizes.

### *6.1.1 Hyper-exponential approximation*

In order to quantify the performance perceived by the users while being served by a CDN, we need to find the P.D.F of waiting time distribution. The Bounded Pareto distribution has all moments finite; however advanced analysis is complex due to the difficulties in manipulating the Laplace transforms of the queuing metrics (e.g. waiting time, busy period). Hence, the 'heavy-tailed' Bounded Pareto distribution can be approximated with a series of exponential distributions (known as Hyper-exponential), while still maintaining the main characteristics of the original distribution, such as heavy tail, first and second moments [7]. An $n$ part Hyper-exponential service distribution has the following P.D.F

$$h_n(t) = \sum_{i=1}^{n} P_i \lambda_i e^{-\lambda_i t}, \text{ where } \sum_{i=1}^{n} P_i = 1$$

which can be used for our purpose.

### 6.1.2 Service distribution and waiting time

The Laplace transform of the service distribution $h_n(t)$ is

$$L_{h_n}(s) = \int_0^{\infty} e^{-st} h_n(t)dt = \sum_{i=1}^{n} \frac{P_i \lambda_i}{\lambda_i + s}$$

The moments of the service distribution can be obtained as

$$E[X^n] = (-1)^n \frac{d^n L_{h_n}(s)}{ds}\bigg|_{s=0}$$

where $X$ is a continuous random variable with P.D.F $h_n(t)$. The first moment (mean) $E[X]$ and the second moment $E[X^2]$ are

$$E[X] = \sum_{i=1}^{n} \frac{P_i}{\lambda_i} \text{ and } E[X^2] = \sum_{i=1}^{n} \frac{2P_i}{\lambda_i^2}$$

The Laplace transform of the waiting time, $L_W(s)$ for an M/G/1 queue with the hyper-exponential approximation of a Bounded Pareto distribution is defined as follows:

$$L_W(s) = \frac{s(1-\rho)}{s - \lambda + \lambda L_{h_n}(s)} = \frac{s(1-\rho)}{s - \lambda + \lambda \sum_{i=1}^{n} \frac{P_i \lambda_i}{\lambda_i + s}}$$

This result can be numerically inverted to obtain the P.D.F of the waiting time distribution, $w(t)$. It can also be used to obtain the cumulative distribution function (C.D.F),

$$W(t) = \Pr[T \le t] = \int_0^t w(t)dt$$

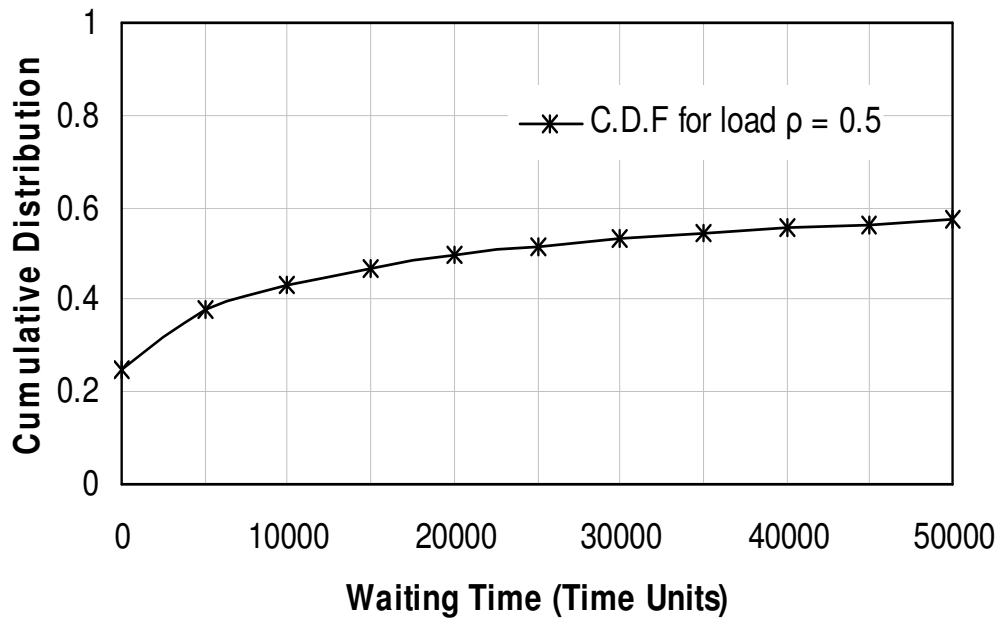

**Figure 10: Cumulative distribution of waiting time of a CDN modeled as an M/G/1 queue**

Using the C.D.F, concrete QoS guarantees can be made regarding the waiting time experienced by a certain percentage of user requests. Figure 10 shows the C.D.F of waiting time of a CDN for the system load $\rho = 0.5$. Here, the hyper-exponential approximation of a Bounded Pareto distribution is used with $\alpha = 1.5$, $k = 1010.15$ and $p = 10^{10}$. From the figure we observe that there is about *50%* probability that the waiting time experienced by the users will be less than *20000* time units. Thus, concrete guarantees can be made regarding the waiting time experienced by a certain percentage of user requests on a given CDN.

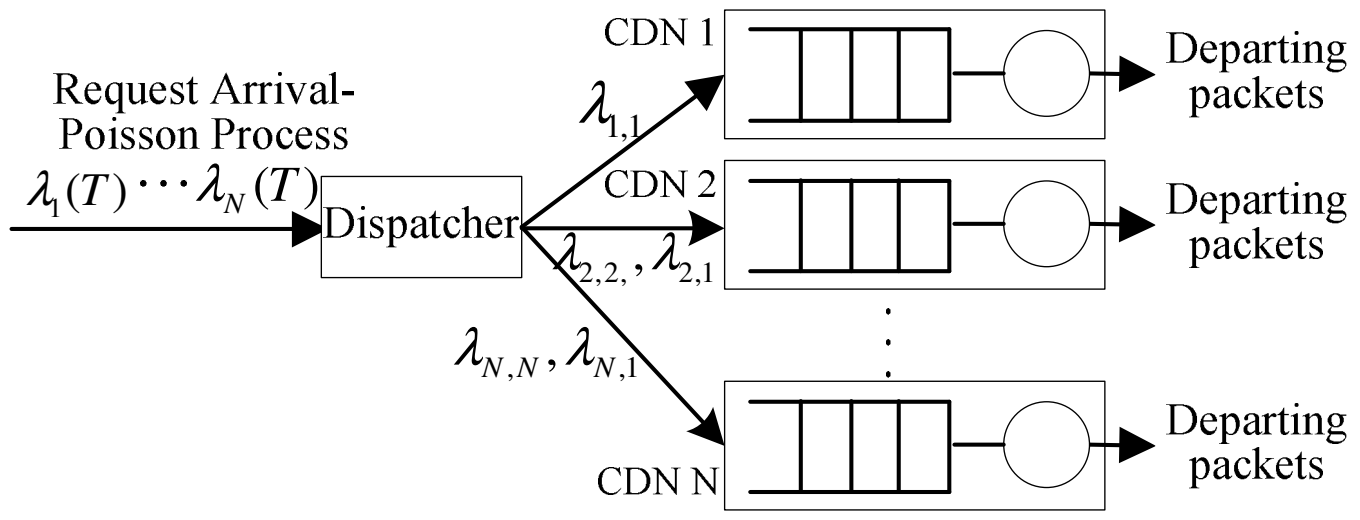

**Figure 11: Conceptual view of the peering CDNs**

## 6.2 Peering CDNs Model

A CDN's inability to meet user QoS requirements according to the SLAs may lead to a collaboration of CDNs, so that it may redirect excess requests to the Web servers of the peers. In Figure 11, a conceptual view of the peering CDNs is provided where each CDN is modeled as an M/G/1 queue. It is abstracted that $N$ independent streams of end-user requests arrive at a conceptual entity, called *dispatcher*, following a Poisson process with the mean arrival rate $\lambda_i$, $i \in \{1,2,\ldots,N\}$. The dispatcher acts as a centralized scheduler in a particular peering relationship with independent mechanism to distribute content requests among partnering CDNs in a user transparent manner. If, on arrival, a user request can not be serviced by CDN $i$, it may redirect excess requests to the peers. Since this dispatching acts on individual requests of Web content, it endeavors to achieve a fine grain control level. We anticipate that it can also pave the ways in performing the *request assignment* and *redirection* at multiple levels – at the DNS, at the gateways to local clusters and also (redirection) between servers in a cluster. The dispatcher follows a certain policy that assists to assign a fraction of requests of CDN *i* to CDN *j*. The request stream to a CDN in the peering CDNs model is defined as $\lambda_{j,i}$ = request to CDN *j* for CDN *i*'s content. For $\forall j \neq i$, $\lambda_{j,i}$ denotes redirected user requests, where CDN *i* is the primary where CDN *j* is a peer. On the other hand, for $\forall j = i$, $\lambda_{j,i}$ denotes the user requests to a primary CDN *i*. For example, request to CDN B for CDN A's content can be denoted as $\lambda_{B,A}$.

The service times of the CDNs are defined as independent of interarrival times and of one another, and have a general P.D.F. Inside each request stream, there is FCFS service. Each request stream is assigned priority. Here, $p = 1,2,\ldots,P$ priority classes of user-requests are assumed. A peer always prioritizes the requests from the primary CDN over its own user requests. However, if a redirected request (higher priority) arrives to a peer when its own user request (lower priority) is being served, it never interrupts the current service. Thus, this priority discipline is non-preemptive during service quantum of end-user requests.

### *6.2.1 Waiting time*

The classical result [12] on non-preemptive head-of-the-line (HOL) priority queue can be used to find the expected waiting time for the *p*-th $(p = 1,2,\ldots,P)$ priority user request,

$$W_p = \frac{W_0}{(1-\sigma_p)(1-\sigma_{p+1})}, \text{ where } \sigma_p = \sum_{i=p}^{P} \rho_i \tag{1}$$

$W_0$ is the average delay to a particular priority user request due to other requests found in service. It can be expressed as,

$$W_0 = \sum_{i=1}^{P} \frac{\rho_i E[X_i^2]}{2}$$

where $E[X_i^2]$ is the second moment of service time for a customer from class *i*. It can be mentioned that $W_k$ does not depend on user requests from lower priority class (i.e. $i = 1, 2,\ldots, p-1$), except for their contribution to the numerator $W_0$ [20].

Let us assume that the user requests for the primary CDN belongs to the *p*-th priority class. The Laplace transform of the waiting time for the primary CDN is denoted as $W^*_p(s)$. Using the known solution [13] for the distribution of waiting time for each priority group in a priority queue, it can be expressed as,

$$W_p^*(s) = \frac{(1-\rho)s + \lambda_L[1 - \sum_{j=1}^{p-1} \frac{\lambda_{j,j}}{\lambda_L} B_j^*(s)]}{s - \lambda_{p,p} + \lambda_{p,p} B_p^*(s)}, \text{ where } \lambda_L = \sum_{j=1}^{p-1} \lambda_{j,j} \tag{2}$$

Similarly, for any peer with the priority in the range *1,2 ...,(p-1)* the Laplace transform of waiting time is found as,

$$W_j^*(s) = \frac{(1-\rho)[s + \lambda_{p,p} - \lambda_{p,p} G_p^*(s)]}{s - \sum_{j=1}^{p-1} \lambda_{j,j} + \sum_{j=1}^{p-1} \lambda_{j,j} \sum_{j=1}^{p-1} B_j^*(s + \lambda_{p,p} - \lambda_{p,p} G_p^*(s))} \tag{3}$$

Here, $G^*_p(s)$ is the transform for the M/G/1 busy period distribution for a *p*-priority class, which is expressed as,

$$G_p^*(s) = B_p^*(s + \lambda_{p,p} - \lambda_{p,p} G_p^*(s)) \tag{4}$$

These solutions can be used to measure the expected waiting time for end-user requests to each of the participating entities in the peering arrangement.

### *6.2.2 QoS performance*

The ability to gauge the QoS of a CDN provider is crucial for achieving effective service from it. The P.D.F of the waiting time distribution (through numerical inversion) for each CDN, with independent priority class can be used to observe the expected waiting time perceived by majority of users in the peering CDNs system. Since a primary CDN's request has priority over any peer's own user requests, we can consider using (2) for a primary CDN, while (3) for any peering CDN. Though these equations are useful for computation, the iterative expression for $G^*_i(s)$ in (4) is impossible to invert numerically. Therefore, the waiting time experienced by a primary CDN's user requests is found using (2), while the classical result presented in (1) is used to find the average expected waiting time for the peer's user requests.

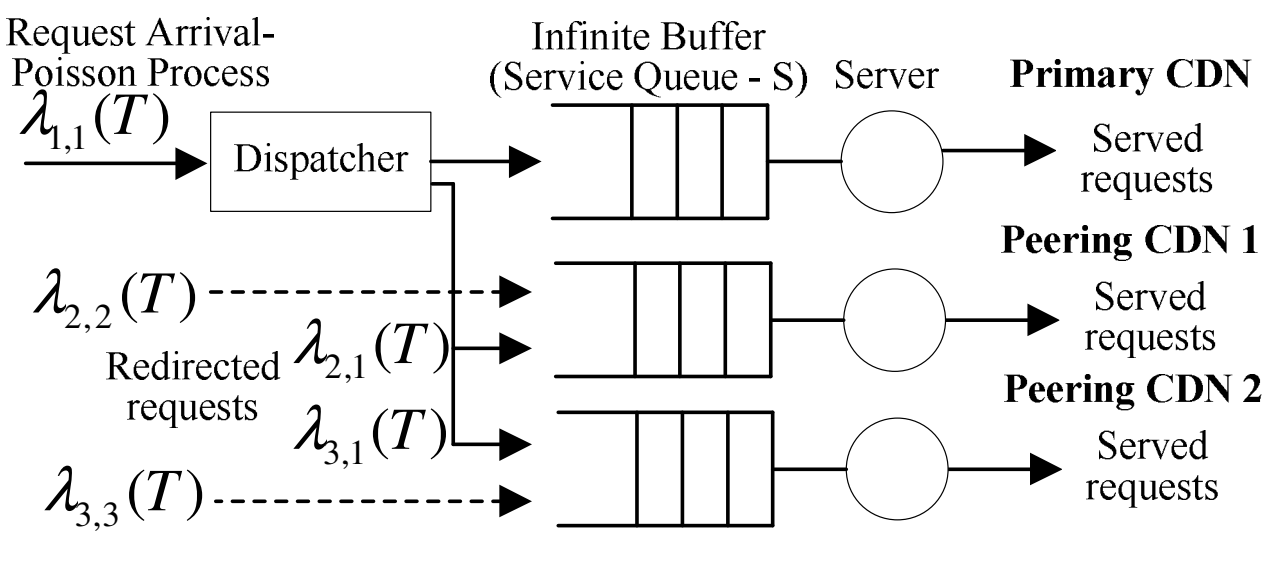

**Figure 12: A peering scenario with three CDNs**

## 7. RESULTS

In this section, we present the performance results obtained using the models presented in Section 6. We consider an established peering CDNs system consisting of three CDNs, as shown in Figure 12. Table 4 lists the notations used for this scenario. Each CDN is modeled as an M/G/1 queue with highly variable Hyper-exponential distribution which approximates a heavy-tailed Bounded Pareto service distribution ($\alpha$, $k$, $p$) with variable task sizes. Thus, the workload model incorporates the high variability and self-similar nature of Web access. Table 5 shows the distributions, probability density functions and parameter ranges for the workload model. For our experiments, we consider the expected waiting time as an important parameter to evaluate the performance of a CDN. In our peering scenario, we also assume that all peers hold the same replicated content and there exists an SLA for serving all user requests by the primary CDN in less than *20000* time units.

**Table 4: List of notations used in the peering CDNs scenario**

| Notation | Description |
|---|---|
| $N$ | Number of CDNs, $N = \{1,2,\ldots,N\}$ |
| $E[X_i^j]$ | $j$-th moment of CDN $i$'s service distribution |
| $\rho_i$ | Initial load on CDN $I$, $\rho_i = \lambda_{i,i}E[X_i]$ |
| $\rho^*_i$ | New load on CDN $i$, $\rho^*_i = \lambda^*_{i,i}E[X_i]$ |
| $\lambda_{i,i}$ | Initial arrival rate at CDN $i$ |
| $\lambda^*_{i,i}$ | New arrival rate at CDN $i$, $\lambda^*_{i,i} = \lambda_{i,i} - \lambda^i_{redirect}$ |
| $E[W_i]$ | Expected waiting time (initial) at primary CDN $i$, |
| $E[W_i^*]$ | Weighted average expected waiting time at CDN $i$, $E[W_i^*] = \frac{\rho - \rho^i_{redirect}}{\rho} E[W_i] + \sum_{k=1}^{i-1} \frac{w_k \cdot \rho^i_{redirect}}{\rho} E[W_k]$ <br> $w_k$ depends on the redirected ratio to a given peer |
| $\rho^i_{redirect}$ | Fraction of content requests redirected from CDN $i$ |

**Table 5: Workload model**

| Category | Distribution | P.D.F | Range | Parameters |
|---|---|---|---|---|
| Primary CDN, $0.1 \leq \rho \leq 0.9$ | Hyper-exponential | $h_n(t) = \sum_{i=1}^{n} P_i \lambda_i e^{-\lambda_i t}$ approximating $f(x) = \frac{\alpha k^\alpha}{1-(k/p)^\alpha} x^{-\alpha-1}$ | $x \geq k$ | $\alpha = 1.5$ <br> $k = 1010.15$ <br> $p = 10^{10}$ |
| Peer 1, $\rho = 0.5$ | Hyper-exponential | $h_n(t) = \sum_{i=1}^{n} P_i \lambda_i e^{-\lambda_i t}$ approximating $f(x) = \frac{\alpha k^\alpha}{1-(k/p)^\alpha} x^{-\alpha-1}$ | $x \geq k$ | $\alpha = 1.5$ <br> $k = 1010.15$ <br> $p = 10^{10}$ |
| Peer 2, $\rho = 0.4$ | Hyper-exponential | $h_n(t) = \sum_{i=1}^{n} P_i \lambda_i e^{-\lambda_i t}$ approximating $f(x) = \frac{\alpha k^\alpha}{1-(k/p)^\alpha} x^{-\alpha-1}$ | $x \geq k$ | $\alpha = 2$ <br> $k = 1500.23$ <br> $p = 10^{10}$ |

## 7.1 QoS performance of the primary CDN

First, we attempt to provide the evidence that a peering arrangement between CDNs is able to assist a primary CDN to provide better QoS to its users. The C.D.F of the waiting time distribution of the primary CDN can be used as the QoS performance metric. In a highly variable system such as peering CDNs it is more significant than average values. The waiting time corresponds to the time elapsed by a user request before being served by the CDN. Figure 13 shows the C.D.F of waiting time of the primary CDN without peering at different loads. From the figure we see that for a fair load $\rho = 0.6$ there is about *55%* probability that users will have a waiting time less than the threshold of *20000* time units. For a moderate load $\rho = 0.7$, there is about *50%* probability for users to have waiting time below the threshold, while for a heavy load $\rho = 0.9$ the probability reduces to > *24%*.

The peering CDNs model arranges the participating providers according to a non-preemptive HOL priority queuing system (Section 6.2). It is an M/G/1 queuing system in which we assume that user priority is known upon their arrival to a CDN and therefore they may be ordered in the queue immediately upon entry. Therefore, various priority classes receive different grades of service and requests are discriminated on the basis of *known* priority. Thus, in our model an incoming request (with priority *p*) joins the queue behind all other user requests with priorities less than or equal to *p* and in front of all the user requests with priority greater than *p*. Due to this nature of the peering CDNs model, the effect of peering can be captured irrespective of any particular request-redirection policy.

Figure 14 shows the C.D.F of the primary CDN with peering for different loads. By comparing Figure 13 and Figure 14, it can be found that for a fair load $\rho = 0.6$ there is about *80%* probability that users will have a waiting time of less than a threshold of *20000* time units. Therefore, peering assists the primary CDN to achieve a QoS performance improvement of about *31%*. For a moderate load $\rho = 0.7$, there is > *81%* probability for users to have waiting time below the threshold, an improvement of about *38%*. For a heavily loaded primary CDN with $\rho = 0.9$ the probability becomes about *70%*, which lead to an improvement of > *65%*. Moreover, for loads $\rho > 0.9$, still higher improvement can be predicted by the model. Based on these observations, it can be stated that peering between CDNs irrespective of any particular request-redirection policy achieves substantial QoS performance improvement when comparing to the non-peering case.

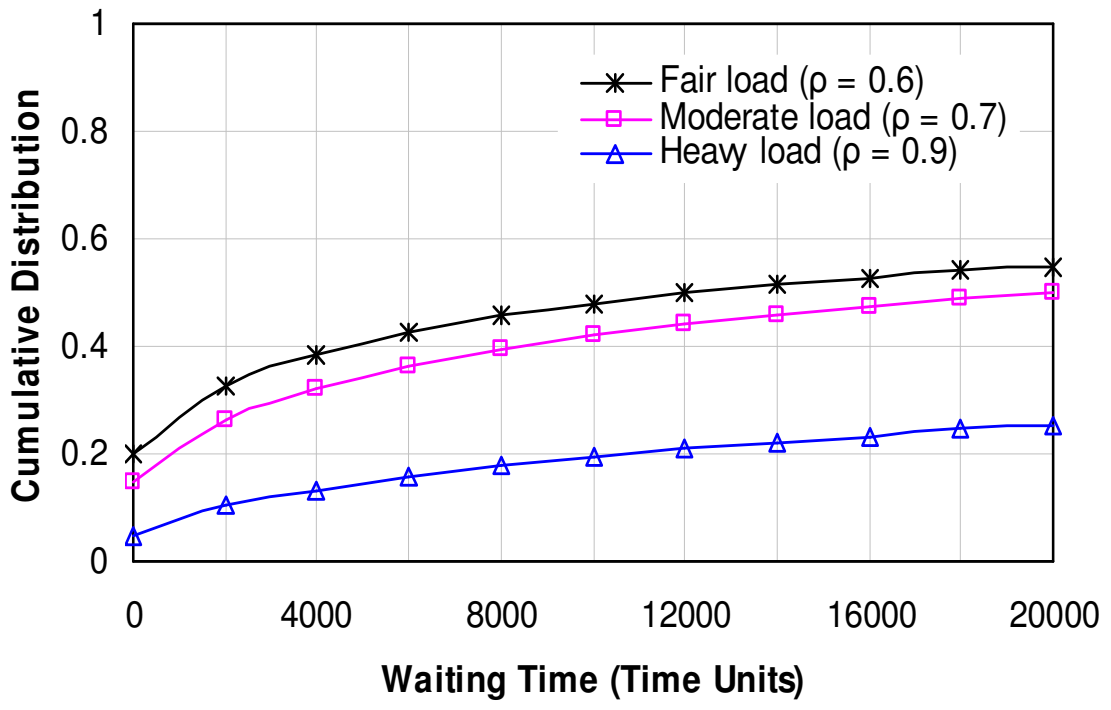

**Figure 13: Cumulative distribution of waiting time of the primary CDN without peering**

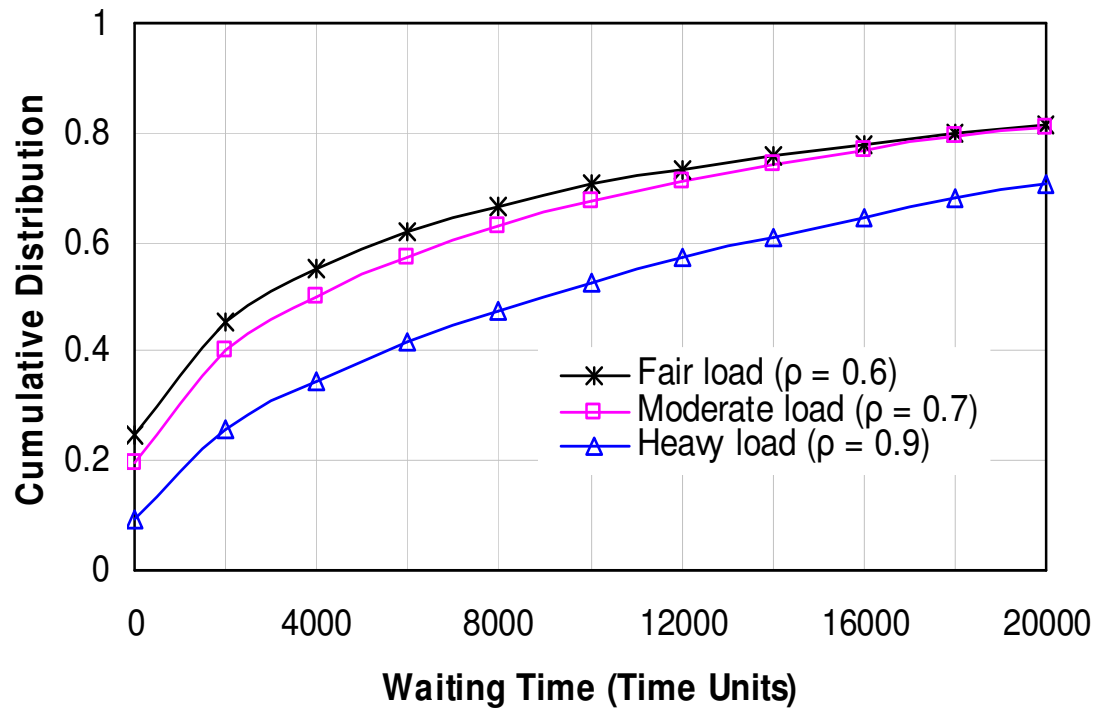

**Figure 14: Cumulative distribution of waiting time of the primary CDN in a peering arrangement**

## 7.2 Request-redirection policies

In the peering CDNs model, no redirection is assumed until primary CDN's load reaches a *threshold* load ($\rho = 0.5$). This load value is also used as the *baseline* load for comparing waiting times at different primary CDN loads. Any load above that will be 'shed' to peers. A request-redirection policy determines which requests have to be redirected to the peers. Each peer is ready to accept only a certain fraction (*acceptance threshold*) of the redirected requests. Any redirected request to a given peer exceeding this acceptance threshold is simply dropped to maintain the system equilibrium. In face of sudden surge in demand, the load on a given primary CDN $i$, $i \in \{1,2,\ldots,N\}$ becomes, $\rho^{*}_{i} = \rho_i - \rho^{i}_{redirect}$ and the redirected load is distributed among the peering CDNs. The value of $\rho^{i}_{redirect}$ varies depending on the dispatcher chosen redirection policy. The initial and new load on the given primary CDN $i$ is measured by, $\rho_i = \lambda_{i,i} E[X_i]$ and $\rho^{*}_{i} = \lambda^{*}_{i,i} E[X_i]$ respectively. $\lambda_{i,i}$ is the initial arrival rate, whereas, $\lambda^{*}_{i,i} = \lambda_{i,i} - \lambda^{i}_{redirect}$ is the new arrival rate.

We define four request-redirection policies for evaluation within the peering CDNs model:

- *Uniform Load Balanced (ULB)* request-redirection policy distributes the redirected content requests uniformly among all the peering CDNs.
- *Minimum Load Balanced (MLB)* request-redirection policy assigns the redirected content requests to the peer with minimum expected waiting time.
- *Probabilistic Load Balanced (PLB)* request-redirection policy distributes redirected content requests to the peers according to certain probability. This probability depends on the load threshold for the fraction of redirected requests that a peer can accept from the primary. In our case, considering a peering system of three CDNs, we use a probability of *0.4* and *0.6* to assign a certain fraction of the redirected requests to peer 1 and peer 2 respectively. We measure this probability based on the acceptance threshold and the service capacity of the peers.
- *Weighted Load Balanced (WLB)* request-redirection policy assigns *80%* of redirected content requests to the peer with minimum expected waiting time. The remaining *20%* of traffic is uniformly distributed over all other participating peers.

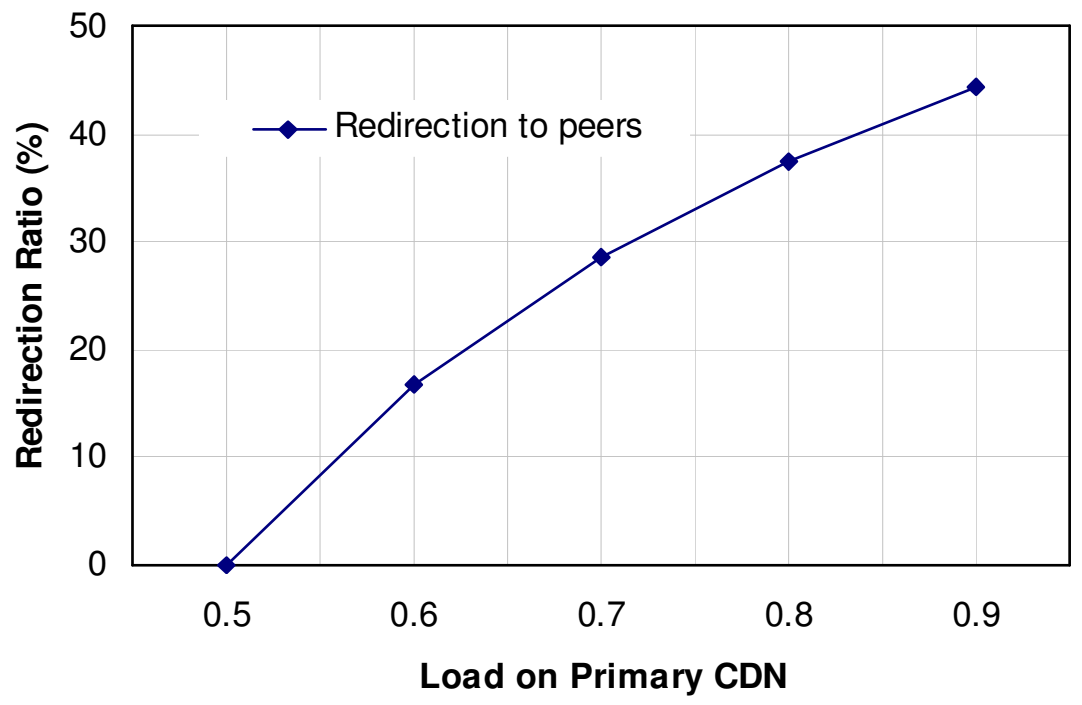

**Figure 15: Redirection ratio at different primary CDN loads**

The amount of redirected requests is denoted as the *redirection ratio*, which is quantified in percent of the primary CDN's load. We express the redirection ratio as a fraction of the primary CDN's load. The influence of different primary CDN loads on the redirection ratio to the peers is shown in Figure 15. From the figure, we can see that the redirection ratio (in percentage, %) increases with increase in the primary CDN's load, independent of any particular request-redirection policy.

The redirected requests are distributed (by the dispatcher) to the peers according to a particular request-redirection policy. In Figure 16, the redirection ratios assigned to the peers are shown. Each curve denotes a different request-redirection policy and x-axis denotes the load on the primary CDN. In ULB request-redirection policy, both peer 1 and peer 2 receive the same amount of redirected requests from the primary CDN. The use of MLB request-redirection policy by the dispatcher assigns all the redirected requests to peer 2, which has the minimum expected waiting time. Hence, no redirected request is assigned to peer 1. If PLB request-redirection policy is used by the dispatcher, it leads to a distribution of *40%-60%* of the redirected requests to peer 1 and peer 2 respectively. A dispatcher following the WLB request-redirection policy assigns *80%* of redirected requests to peer 2 (with minimum expected waiting time) and rest *20%* to peer 1.

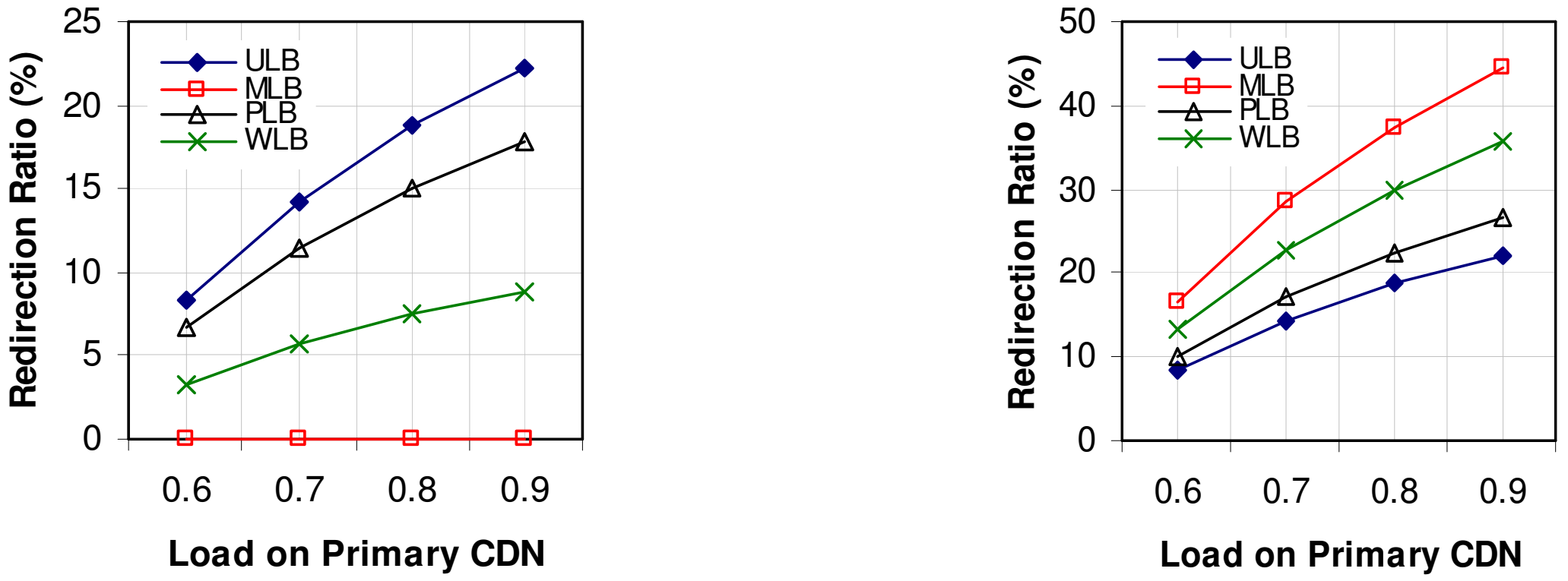

(a) Peer 1 (b) Peer 2

**Figure 16: Distribution of redirected requests to the peers for different request-redirection policies and primary CDN loads**

## 7.3 Impact of request redirection

Next, we study the impact of request-redirection on the expected waiting time of users on the primary CDN for different request-redirection policies. Without request-redirection when the primary CDN's load approaches to *1.0*, the user perceived performance (in terms of waiting time and queue length) for service by the primary CDN tends to infinity. On the other hand, with request-redirection the waiting time of the primary CDN decreases as the requests are redirected to the peers. However, request-redirection may lead to temporary overload on certain peer(s).

Figure 17 shows the expected waiting time experienced by the redirected requests on the peering CDNs for different request-redirection policies. From the figure it can be seen that as more requests are redirected to the peers they realize higher waiting time due to the peers' own load.

Typically a burst of redirected requests improves performance on the primary CDN. In Figure 17, we present this evidence by showing the performance improvement (in terms of waiting time) the primary CDN gains for all the request-redirection policies. Here, we compare the expected waiting time as a function of system load under the four request-redirection policies, by considering lightly loaded peers (load of peer 1 and peer 2 are set to $\rho = 0.5$ and $\rho = 0.4$ respectively), while tuning the primary CDN's load ($0.1 \leq \rho \leq 0.9$). It can be noted that a weighted average value of waiting time is presented in order to capture the effect of request-redirection.

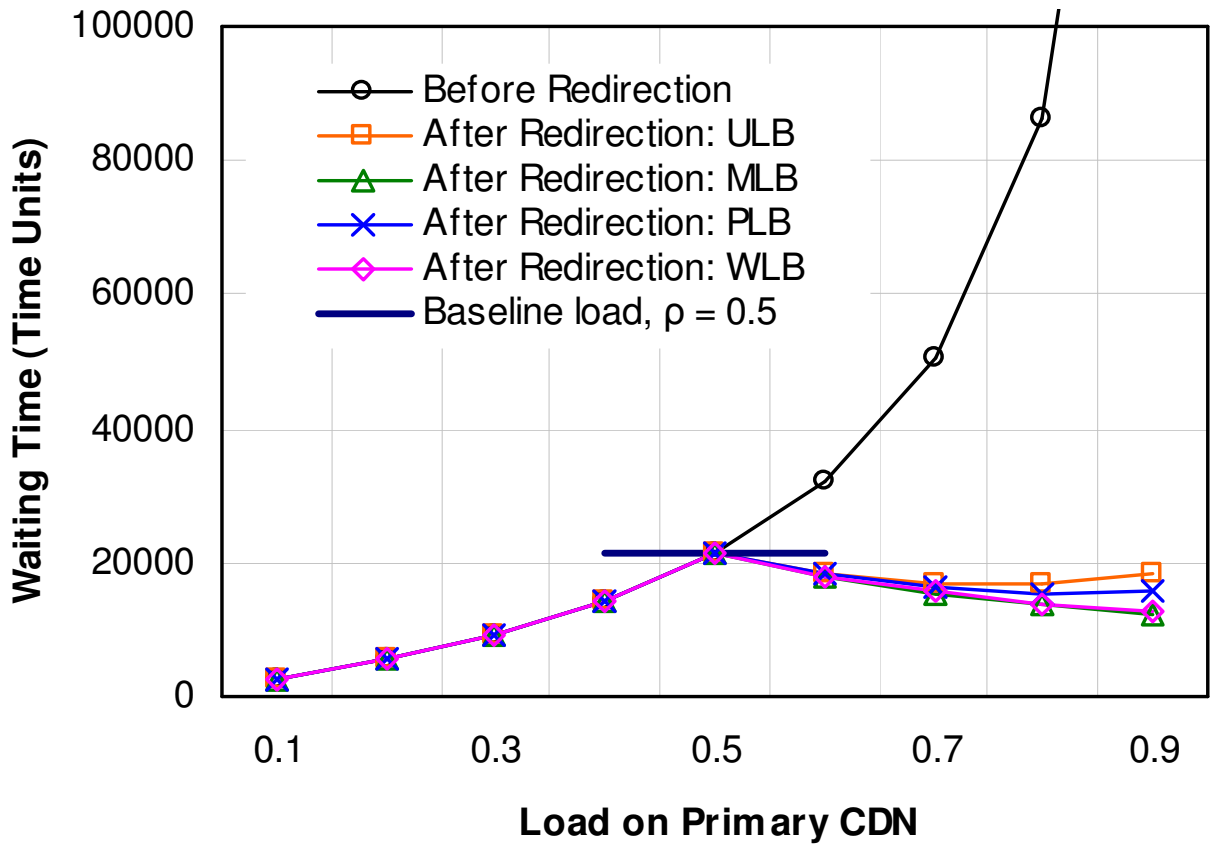

**Figure 17: Impact of request-redirection on waiting time of the primary CDN for different request-redirection policies**

**Table 6: Reduction on waiting time for the primary CDN under different request-redirection policies**

| **Load on primary CDN** | **Reduction in waiting time %** | | | |
|---|---|---|---|---|
| | **ULB** | **MLB** | **PLB** | **WLB** |
| Fair load, $\rho = 0.6$ | *43.20%* | *44.41%* | *43.66%* | *44.24%* |
| Moderate load, $\rho = 0.7$ | *66.31%* | *69.31%* | *67.50%* | *68.91%* |
| Heavy load, $\rho = 0.9$ | *90.52%* | *93.70%* | *91.94%* | *93.39%* |

Table 6 summarizes the reduction on the expected waiting time for the primary CDN in peering for different request-redirection policies. For all the four policies, it is also observed that substantial performance improvement is achieved on the expected waiting time when compared to the non-peering case. Among all the four policies, ULB, PLB and WLB effectively have *redirection ratio* as the common performance parameter. For all these three policies, redirected requests are distributed among peers according to certain percentage. Therefore, to some extent they exhibit similar characteristics. Whereas MLB assigns all redirected requests to a single peer. Though results for MLB may show as good performance as the other three policies (due to light load on peers), there is a possible concern for the peer with minimum expected waiting time to become overloaded with the redirected requests (herd effect [16]). Therefore, it is preferable to spread the load of redirected requests among multiple CDNs rather than assigning all redirected requests to a single peer.

From the results, it is clear that all the request-redirection policies guarantee that the maximum waiting time is below *20000* time units. This confirms that redirecting only a certain fraction of requests reduces instability and overload in the system because the peers are not overwhelmed by bursts of additional requests.

## 7.4 Measurement errors

The dispatcher bases its redirection decision on the measured value of the primary CDN's load. So far we have assumed that perfect information is available for this decision. However, the dispatcher can have inaccurate information about the load on the primary CDN e.g. due to delays in receiving the measurements. Therefore, the impact of measurement errors on the effectiveness of the redirection policies can be measured. Let us denote the measured load of the primary CDN at the dispatcher by $\tilde{\rho} = \tilde{\lambda} E[X]$, with $\tilde{\lambda} = \lambda(1 \pm \varepsilon)$, where $\varepsilon$ is the percentage of the correct load $\rho$.

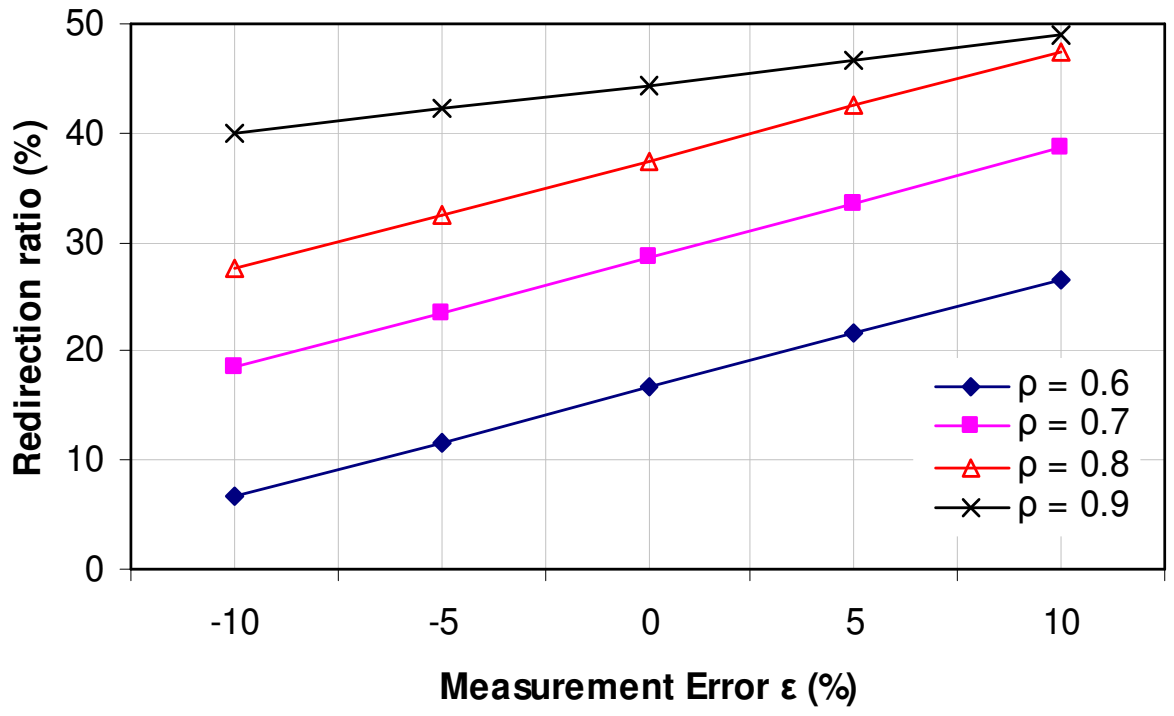

**Figure 18: Impact of measurement errors on request-redirection ratio at different primary CDN load**

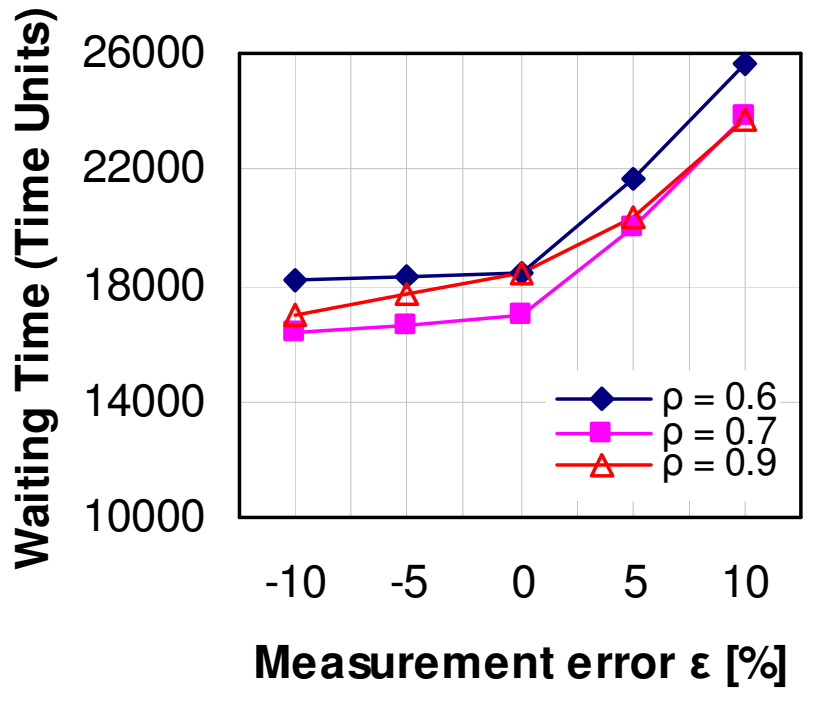

(a) ULB

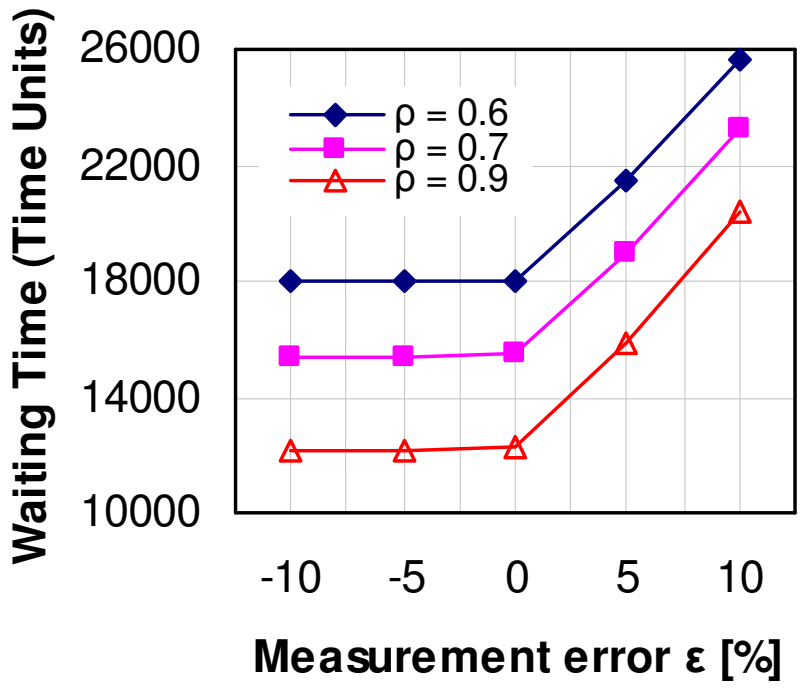

(b) MLB

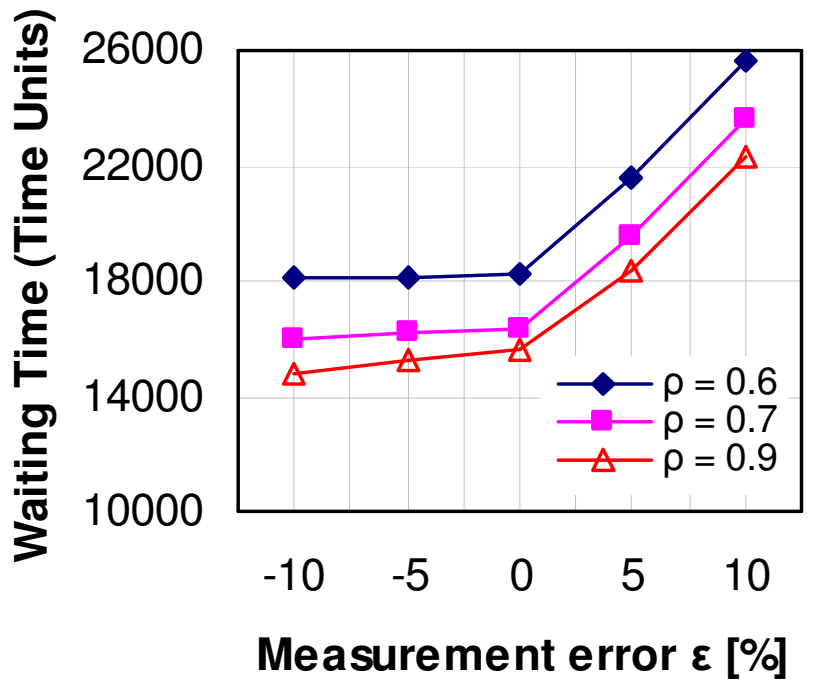

(c) PLB

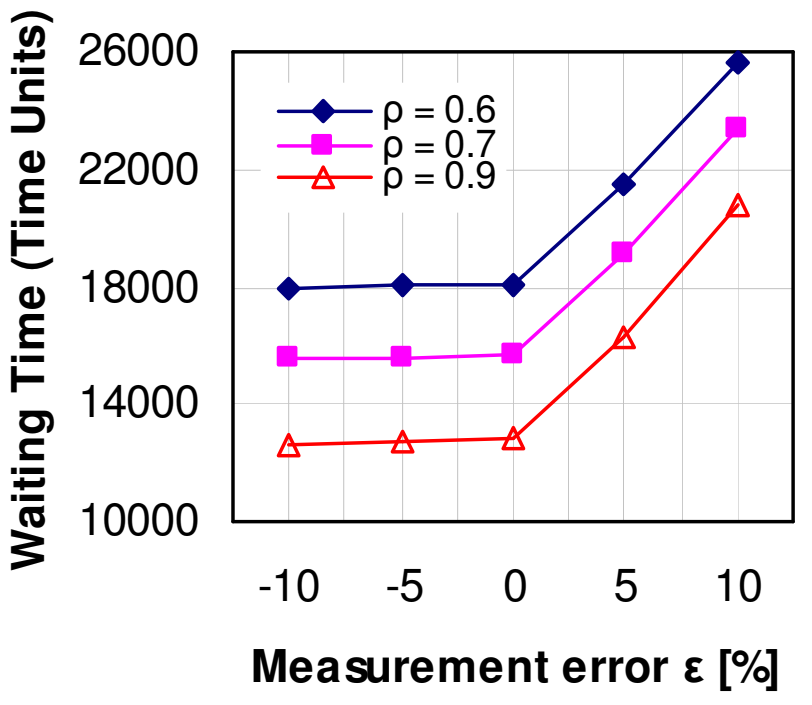

(d) WLB

**Figure 19: Waiting time for load measurement errors $\varepsilon$ for different request-redirection policies**

The effects of primary CDN's load measurement error on the redirection ratio are shown in Figure 18. Each line in the figure denotes a different primary CDN load $\rho$, and the x-axis denotes the measurement error $\varepsilon$, in percent of $\rho$. A value of *0* on the x-axis corresponds to perfect primary CDN load information. From the figure we see that the redirection ratio changes more for positive $\varepsilon$ than for the corresponding negative $\varepsilon$.

Figure 19 shows the impact of primary CDN's load measurement error on the waiting time for different request-redirection policies. It can be observed that in all the four cases, for measurement error $\varepsilon > 0$, the dispatcher assumes the primary CDN's load to be higher than what it is and hence it redirects more requests than the actual load. The extra redirections incur additional waiting time for the user requests and hence it increases linearly from $\varepsilon = 0$. For negative $\varepsilon$, the dispatcher assumes the primary CDN's load to be less than the actual and hence redirects pessimistically. As a result, requests on the primary CDN could experience greater expected waiting time for being processed. However, the average of waiting time normalizes the adverse effect in order to keep the performance at an acceptable level. Nevertheless it can be concluded that greater accuracy is needed in load measurement of the primary CDN.

# 8. CRITICAL EVALUATION

Although our approach can be assistive for peering between CDNs, there are a number of challenges, both technical and non-technical (i.e. commercial and legal), that could hinder its rapid growth. These challenges must be dealt to promote CDN peering. For CDNs to peer, they need a common protocol to define the technical details of their interaction as well as the duration and QoS expected during the peering period. The proprietary nature of a CDN to gain competitive advantage in the market may block off the nascence of peering CDNs in commercial domain. Furthermore, there can often be complex legal issues involved (e.g. embargoed or copyrighted content) that could prevent CDNs from arbitrarily cooperating with each other. Finally, there may simply be no compelling commercial reason for a large CDN provider such as Akamai to participate in CDN peering, given the competitive advantage that its network has the most pervasive geographical coverage of any commercial CDN provider. However, our approach can be beneficial and applicable in research-based academic CDN domain where the main focus is not on whether such peering *will* emerge in reality, which mostly depends on the key players in commercial domain that we cannot divine, but rather on whether such peering *could* emerge.

Although the performance models in this context are simplified in order to accommodate the system complexities, we believe that our models provide a foundation for performing effective peering between CDNs though achieving target QoS in service delivery to end-users. Since the peering CDNs retain load-balancing control within their own Web server sets, using our approach a primary CDN can realize the QoS performance it can provide to the end-users, without requiring individual partners to provide expected service performance from it. Our model-based approach is important since having each CDN provider communicate how it would service millions of potential end-users would introduce significant scalability issues, and requesting this information from each partnering provider at the user requests time would introduce substantial delays. Thus, we believe that our approach seeks to achieve scalability for a CDN in a user transparent manner.

# 9. CONCLUSIONS AND FUTURE WORK

In this paper, we present an open and scalable system to assist the creation of open content delivery networks. In our architecture, when the load on the primary CDN exceeds its capacity, it peers with other CDNs, and the excess end-user requests are offloaded to the Web servers of the peering CDNs. Our contribution lies in designing an architecture for the VO-based peering approach that endeavors to reduce setup and maintenance cost of network infrastructures, while also respecting end-user performance requirements through proper policy management for negotiated SLAs. In addition, it promotes extended scalability and resource sharing with other CDNs through cooperation and coordination. We have also proposed an innovative approach to model the peering CDNs. Through the presented performance models we demonstrated the effects of peering and predicted end-user perceived performance from a primary CDN. We outlined a measurement-based methodology which endeavors to assist in making concrete QoS guarantees by a CDN provider. Our approach assists an overloaded CDN to immediately stabilized by offloaded a fraction of the incoming content requests to the peers.

No prior work done in the content internetworking domain has considered VO-based peering among CDNs. previous research efforts Moreover, many of the related research efforts in this context suggest that only modest progress has been made to define necesssary frameworks and policies for CDN peering. In addition, many of them make strong assumptions on the characteristics of applications without virtualzing multiple providers for cooperative management and delivery of content in a peering environment. Moreover, these systems have not explored the issue of policy management.

Our future work includes using market models in this context in order to encourage resource sharing and peering among different CDNs at global level. This approach is inspired by the successful utilization of economic concepts in management of autonomous resources in global grids [9]. The use of economic concepts in this context would provide a solid basis for rational agents to decide whether to join in peering arrangements. The use of economic models may be the basis for a dynamic replication mechanism that makes replication decisions to utilize surrogates in areas which exhibit the potential to generate hotspots. Our initial work on this issue can be found in [27]. Our future work also includes performing an advanced system analysis to study the impact of other performance parameters such as network latency and cost of peering. It also includes developing a proof-of-the-concept implementation for demonstrating the real-time application of our approach for peering between CDNs.

The methodologies presented in this paper have the potential to motivate and direct further research in finding best practice techniques in measuring and disseminating load information, performing request assignment and redirection, and enabling content replication for CDNs participating in VO-based peering. Our approach endeavors to improve scalability and resource sharing between CDNs through peering, thus evolving past the current landscape where disparate CDNs exist. We expect that our approach for forming VO-based peering CDNs and performance modeling to predict user perceived performance will be a timely contribution to the ongoing content-networking trend. For more information about our efforts on peering CDNs, please visit the project Web site at www.gridbus.org/cdn.

# ACKNOWLEDGEMENTS

We are thankful to James Broberg (The University of Melbourne, Australia), Kyong Hoon Kim (Gyeongsang National University, South Korea) and Kris Bubendorfer (Victoria University of Wellington, New Zealand) for sharing thoughts on this topic, and for discussion regarding the formulation of the system model. We also thank anonymous reviewers for their valuable and constructive comments to improve the paper's content, structure, and readability.